\documentclass[fleqn,usenatbib]{mnras}

\usepackage{newtxtext,newtxmath}

\usepackage[T1]{fontenc}

\DeclareRobustCommand{\VAN}[3]{#2}
\let\VANthebibliography\thebibliography
\def\thebibliography{\DeclareRobustCommand{\VAN}[3]{##3}\VANthebibliography}

\usepackage{graphicx}	
\usepackage{amsmath}	
\usepackage{multirow}
\usepackage{multicol}
\usepackage[nameinlink,noabbrev,capitalize]{cleveref}

\newcommand{\tess}{\textit{TESS}}
\newcommand{\swift}{\textit{Swift}}

\newcommand{\gp}{g^\prime}
\newcommand{\rp}{r^\prime}

\newcommand{\napier}{\mathrm{e}}
\newcommand{\ebv}{E(B-V)}
\newcommand{\vsini}{\mathrm{v}\sin{i}}
\newcommand{\Teff}{{T_\mathrm{eff}}}
\newcommand{\Rsun}{{R_{\sun}}}
\newcommand{\Msun}{{M_{\sun}}}
\newcommand{\Rstar}{{R_*}}
\newcommand{\Mstar}{{M_*}}
\newcommand{\mj}{{m_J}}
\newcommand{\mbbj}{{m^\mathrm{bb}_J}}
\newcommand{\vcrit}{{\mathrm{v}_\mathrm{b}}}
\newcommand{\nub}{{\nu_\mathrm{b}}}

\newcommand{\Nphoton}{\mathcal{N}}
\newcommand{\Npass}{{\Nphoton_\mathrm{pass}}}
\newcommand{\Nscat}{{\Nphoton_\mathrm{scatt}}}
\newcommand{\Nret}{{\Nphoton_\mathrm{ret}}}

\newcommand{\kmps}{\mathrm{km\,s^{-1}}}
\newcommand{\dinv}{\mathrm{d^{-1}}}
\newcommand{\cps}{\mathrm{photons\,s^{-1}\,cm^{-2}}}
\newcommand{\eps}{\mathrm{erg\,s^{-1}}}
\newcommand{\gpqcm}{\mathrm{g\,cm^{-3}}}

\newcommand{\brkts}[1]{\left(#1\right)}
\newcommand{\brktm}[1]{\left\{#1\right\}}
\newcommand{\brktl}[1]{\left[#1\right]}
\newcommand{\diff}[1]{{\mathrm{d}#1}}

\title[
    Pulsation \& disk-growth anti-correlations
]{
    Possible anti-correlations between pulsation amplitudes\\
    and the disk growth of Be stars in giant-outbursting Be X-ray binaries
}

\author[M. Niwano et al.]{
    Masafumi Niwano,$^{1}$\thanks{E-mail: niwano@hp.phys.titech.ac.jp}
    Michael M. Fausnaugh,$^{2}$
    Ryan M. Lau,$^{3}$
    Kishalay De,$^{4}$
    Roberto Soria,$^{5,6,7}$
    \newauthor
    George R. Ricker,$^{4}$
    Roland Vanderspek,$^{4}$
    Michael C. B. Ashley,$^{8}$
    Nicholas Earley,$^{9}$
    \newauthor
    Matthew J. Hankins,$^{10}$
    Mansi M. Kasliwal,$^{9}$
    Anna M. Moore,$^{11}$
    Jamie Soon,$^{11}$
    Tony Travouillon,$^{11}$
    \newauthor
    Mahito Sasada,$^{12}$
    Ichiro Takahashi,$^{1}$
    Yoichi Yatsu,$^{1}$
    and Nobuyuki Kawai$^{1}$
    \\$^{1}$Department of Physics, Tokyo Institute of Technology, 2-12-1 Ookayama, Meguro-ku, Tokyo 152-8551, Japan
    \\$^{2}$Department of Physics \& Astronomy, Texas Tech University, Lubbock, TX 79410-1051, USA
    \\$^{3}$NSF's NOIRLab, 950 N. Cherry Avenue, Tucson, AZ 85719, USA
    \\$^{4}$MIT-Kavli Institute for Astrophysics and Space Research, 77 Massachusetts Avenue, Cambridge, MA 02139, USA
    \\$^{5}$College of Astronomy and Space Sciences, University of the Chinese Academy of Sciences, Beijing 100049, China
    \\$^{6}$INAF-Osservatorio Astrofisico di Torino, Strada Osservatorio 20, I-10025 Pino Torinese, Italy
    \\$^{7}$Sydney Institute for Astronomy, School of Physics A28, The University of Sydney, Sydney, NSW 2006, Australia
    \\$^{8}$School of Physics, University of New South Wales, Sydney NSW 2052, Australia
    \\$^{9}$Cahill Center for Astrophysics, California Institute of Technology, 1200 E. California Blvd. Pasadena, CA 91125, USA
    \\$^{10}$Arkansas Tech University, Russellville, AR 72801, USA
    \\$^{11}$Research School of Astronomy and Astrophysics, Australian National University, Canberra, ACT 2611, Australia
    \\$^{12}$Institute of Innovative Research, Tokyo Institute of Technology, 2-12-1 Ookayama, Meguro-ku, Tokyo 152-8551, Japan
}

\date{Accepted XXX. Received YYY; in original form ZZZ}

\pubyear{2024}

\begin{document}
\label{firstpage}
\pagerange{\pageref{firstpage}--\pageref{lastpage}}
\maketitle

\begin{abstract}
The mechanism of X-ray outbursts in Be X-ray binaries remains a mystery, and understanding their circumstellar disks is crucial for a solution of the mass-transfer problem.
In particular, it is important to identify the Be star activities (e.g., pulsations) that cause mass ejection and, hence, disk formation.
Therefore, we investigated the relationship between optical flux oscillations and the infrared (IR) excess in a sample of five Be X-ray binaries.
Applying the Lomb-Scargle technique to high-cadence optical light curves from the Transiting Exoplanet Survey Satellite ($\tess$), we detected several significant oscillation modes in the 3 to 24 hour period range for each source.
We also measured the IR excess (a proxy for disk growth) of those five sources, using J-band light curves from Palomar Gattini-IR.
In four of the five sources, we found anti-correlations between the IR excess and the amplitude of the main flux oscillation modes.
This result is inconsistent with the conventional idea that non-radial pulsations drive mass ejections.
We propose an alternative scenario where internal temperature variations in the Be star cause transitions between pulsation-active and mass-ejection-active states.
\end{abstract}

\begin{keywords}
stars: early-type -- stars: emission-line, Be -- stars: oscillations -- stars:rotation -- binaries: close -- X-rays:binaries
\end{keywords}



\section{Introduction}\label{sec:intro}
A Be X-ray binary (BeXB) is a binary system composed of a compact object (a neutron star in most cases) and a Be star, an early type star with a circumstellar disk \citep{reig_2011}.
In addition to standard observational features of Be stars such as double-peaked emission lines and infrared (IR) excess (the latter usually explained as thermal bremsstrahlung from the disk), BeXBs exhibit various peculiar phenomena caused by the interaction of the compact object with the circumstellar disk.
The most notable of them are normal (Type-I) and giant (Type-II) outbursts.
Normal outbursts occur regularly and (quasi-) periodically, and are caused by the compact object capturing mass from the circumstellar disk when passing around the periastron.
Their typical duration and luminosity are about 0.2--0.3 times the orbital period and $\lesssim 10^{37}~\eps$, respectively.
On the other hand, giant outbursts are rarer, more luminous and unpredictable; their mechanism is not well understood.
They last several times the orbital period, and their luminosity often exceeds $\sim10^{38}~\eps$ \citep{reig_2013}, the Eddington limit of a neutron star.
In at least three well-studied cases (SMC X-3: \citealp{tsygankov_2017}, \citealp{townsend_2017}; Swift J0243.6+6124: \citealp{wilson_2018}, \citealp{alfonso_2024}; RX J0209.6-7427: \citealp{vasilopoulos_2020}, \citealp{hou_2022}, \citealp{west_2024}), the peak X-ray luminosity reached $\approx2\times10^{39}~\eps$, placing those BeXBs into the ultra-luminous X-ray (ULX) pulsar class.
Therefore, giant-outbursting BeXBs might be a significant component of the observed ULX population in the local universe \citep{eanshaw_2018,kuranov_2020,gurpide_2021,misra_2024}.
In particular, luminous BeXBs can be the dominant component of the X-ray binary population in galaxies where star formation has peaked $\approx25\text{--}60$ Myr ago \citep{antoniou_2010,antoniou_2016}, as opposed to more recent (dominated by high-mass X-ray binaries) or older (dominated by low-mass X-ray binaries) star-forming environments.
Thus, understanding the connection between donor-star variability and giant outbursts in BeXBs is an important step for the modelling of X-ray binary populations in external galaxies.
Such studies also provide observational constraints to theoretical model of super-critical accretion onto magnetised neutron stars \citep{tsygankov_2017,mushtukov_2017,mushtukov_2019}.

Several studies \citep{martin_2011,martin_2014,okazaki_2013,reig_2018} suggest that giant outbursts happen when the outer radius of the circumstellar disk (possibly warped, inclined or eccentric) reaches the orbit of the compact object and the latter captures a large amount of mass in a short time.
A circumstellar disk is formed via outwards viscous diffusion of mass ejected from the stellar surface \citep{rivinius_2013,rimulo_2018}.
However, the mass ejection mechanism is still unclear.
Although Be stars are generally rapid rotators, their rotation velocities are typically only about 70 per cent of their critical velocity \citep{rivinius_2013}.
This means that their rotation alone cannot explain mass ejections.
Non-radial pulsations are an alternative mechanism proposed to explain mass ejections \citep{bagnulo_2012}.
In support of this scenario, observations of the B2Ve star $\mu$ Cen showed a coincidence of mass ejections with constructive interference of the pulsational velocity fields \citep{rivinius_1998}.
However, theoretically, non-radial pulsations limited by the speed of sound seem to be insufficient for disk formation \citep{smith_2012,torrejon_2012}.
Addressing this mass ejection problem is the main objective of this work.

The operational launch of the Transiting Exoplanet Survey Satellite ($\tess$; \citealp{ricker_2015}) in 2018 was a notable event for the asteroseismology of massive stars, including Be stars.
Although the main purpose of $\tess$ is to explore exoplanets, its high-precision and high-cadence continuous light curves are very useful for the analysis of periodic flux oscillations of massive stars due to their pulsations and rotation.
Several $\tess$ studies of Be stars and BeXBs have already shed new light in this field.
\citet{balona_2020} examined $\tess$ light curves of 57 Be stars, and found that 74 per cent of their targets show a single or harmonic double peak in the 0.5--3 $\dinv$ frequency range of their power spectra.
They argued that such peaks are caused by stellar rotation.
\citet{ramsay_2022} investigated $\tess$ light curves of 23 high-mass X-ray binaries, and confirmed the detection of 0.1--1 $\dinv$ oscillations in the power spectra of all their targets.
\citet{reig_2022} used $\tess$ to study the power spectrum morphology of 22 BeXBs, and found that BeXBs and classical Be stars are indistinguishable in terms of pulsational characteristics.

In this work, we analyse the $\tess$ light curves of BeXBs to investigate long-term variations in the activity level of Be donor stars.
We compare the results with the long-term multi-wavelength light curves in order to constrain the mass ejection mechanism of the Be donor star.
This article is composed as follows.
In \cref{sec:target}, we provide a brief description of the target selection.
\cref{sec:observation} describes the observational data used in this work: $\tess$, $\swift$, MAXI, ZTF, ATLAS, and Gattini-IR.
We explain our Lomb-Scargle analysis of the $\tess$ light curves and the IR excess evaluation in \cref{sec:analysis}.
\cref{sec:result} shows long-term multi-wavelength light curves and amplitude spectra of the target sources.
We highlight the discovery of an anti-correlation between the IR excess and the amplitude of the main flux oscillation modes.
In \cref{sec:discussion}, we discuss possible interpretations of this observed anti-correlation.
We propose that internal temperature variations in the Be star cause transitions between pulsation-active and mass-ejection-active states.
Finally, \cref{sec:conclusion} concludes this article.
\section{Target selection}\label{sec:target}
For BeXBs which have an optical magnitude of $\lesssim14~\mathrm{mag}$ and available $\tess$ light curves, and are in the northern sky, we checked multi-wavelength light curves (cf. \cref{sec:lightcurves}) in the modified Julian date (MJD) range of 58400--60000 and found that five of them showed significant variations in X-ray, optical and near-infrared (NIR) band (cf. \cref{sec:lc_result}).
In most cases, optical and infrared (OIR) variations of Be stars are derived by the growth or the decay of their circumstellar disks.
Thus we selected these five BeXBs as the targets for this study.
\cref{tab:target_list} describes the selected five sources: 4U0115+634, GRO J2058+42, RX J0440.9+4421, SAX J2103.5+4545, and V0332+53.

These are already known to exhibit long-term multi-wavelength activities such as giant outbursts, disk-derived OIR variations.
For 4U0115+634, long-term variations have been studied in \citet{negueruela_2001} and \citet{reig_2007}, and they reported quasi-periodic disk variations with an approximately 5-year cycle and giant outbursts linked to them.
Furthermore, variations in polarization were observed that appear to be related to disturbances in the disk during the giant outburst \citep{reig_2018}.
For GRO J2058+42, \citet{reig_2023} confirmed sinusoidal variations in optical light curves and the H$\alpha$ equivalent width with a period of about 9.5 years and occurrences of giant outbursts related to them, and stated that the symmetry of the double-peaked emission lines suggested that the disk was not warped.
\citet{reig_2005} reported long-term variability in OIR light curves and H$\alpha$ equivalent width of RX J0440.9+4431, and suggested that disturbances in the disk due to interactions with neutron stars can only occur if the disk is sufficiently developed, based on V/R\footnote{Intensity ratio of double-peaked emission lines} variations.
Long-term activities of SAX J2103.5+4545 are studied in \citet{reig_2010} and \citet{camero_2014}.
This source shows highly correlated X-ray and OIR variability, and X-ray flares linked to peaks of OIR light curves.
One notable feature is that the orbital period of this source ($P_\mathrm{orb}=12.67~\mathrm{d}$; \citealp{camero_2007}) is particularly short for known BeXBs.
\citet{caballero_2016} carried out a long-term X-ray and OIR observation of V0332+53, and suggested that the inner regions surrounding the magnetosphere was visualized during the lowest flux state based on the presence of X-ray quasi-periodic oscillation at that time.

\begin{table*}
    \caption{
        Target BeXBs.
    }
    \begin{tabular}{ccccccccccc}\hline
        Name & R.A.$^\mathrm{[a]}$ & Dec.$^\mathrm{[a]}$ & SpT$^\mathrm{[a]}$ & \multicolumn{3}{c}{Magnitude [AB]} & Distance$^*$ & $P_\mathrm{orb}$ & $\vsini$ & $\ebv^\mathrm{[c]}$\\
         & [H:M:S] & [D:M:S] & & $V^\mathrm{[a]}$ & $\tess$$^\mathrm{[b]}$ & $J^\mathrm{[a]}$ & [kpc] & [d] & [$\kmps$] & \\\hline
        4U0115+634 & 01:18:32 & 63:44:33 & B0.2Ve & 15.3 & 13.3 & 12.5 & $7.3\pm0.9$ & $24.3^\mathrm{[1a]}$ & $\approx300^\mathrm{[1b]}$ & $1.71\pm0.05$\\
        GRO J2058+42 & 20:58:48 & 41:46:37 & O9.5-B0IV-Ve & 14.9 & 13.3 & 12.7 & $12.9\pm2.5$ & $55.03^\mathrm{[2a]}$ & $\approx250^\mathrm{[2b]}$ & $1.37\pm0.03$\\
        RX J0440.9+4431 & 04:41:00 & 44:31:49 & B0.2Ve & 10.7 & 9.9 & 10.4 & $2.6\pm0.1$ & $150^\mathrm{[3a]}$ & $235\pm15^\mathrm{[3b]}$ & $0.91\pm0.03$\\
        SAX J2103.5+4545 & 21:03:36 & 45:45:06 & B0Ve & 14.3 & 13.0 & 12.8 & $7.6\pm0.8$ & $12.67^\mathrm{[4a]}$ & $240\pm20^\mathrm{[4b]}$ & $1.36\pm0.03$\\
        V0332+53 & 03:35:00 & 53:10:23 & O8.5Ve & 15.5 & 13.0 & 12.7 & $7.4\pm1.1$ & $34.25^\mathrm{[5a]}$ & $\approx150^\mathrm{[5a]}$ & $1.94\pm0.03$\\
        \hline
        \multicolumn{11}{l}{
            [a] \citet{wegner_2000}; [b] \citet{paegert_2022}; [c] \citet{reig_2015}; [1a] \citet{tamura_1992}; [1b] \citet{negueruela_2001};
        }\\
        \multicolumn{11}{l}{
             [2a] \citet{wilson_2005}; [2b] \citet{kizilovglu_2007}; [3a] \citet{ferrigno_2013}; [3b] \citet{reig_2005}; [4a] \citet{camero_2007}; 
        }\\
        \multicolumn{11}{l}{
            [4b] \citet{reig_2004}; [5a] \citet{negueruela_1999}; $^*$ Distances are calculated from Gaia DR3 parallaxes \citep{gaia_2023}.
        }
    \end{tabular}
    \label{tab:target_list}
\end{table*}
\section{Observations}\label{sec:observation}
We utilised high-cadence optical light curves of $\tess$, and long-term multi-wavelength light curves.

\subsection{Transiting Exoplanet Survey Satellite}\label{sec:tess}
Transiting Exoplanet Survey Satellite ($\tess$; \citealp{ricker_2015}) is a optical space telescope operated by National Aeronautics and Space Administration (NASA) and Massachusetts Institute of Technology (MIT), and its main purpose is to explore exoplanets by observing their transits.
Full frame images (FFIs) at 10 or 30-minutes intervals are available from the Mikulski Archive for Space Telescopes (MAST).

We carried out the photometry of FFIs with the pipeline of $\tess$ Transients project \citep{fausnaugh_2021,fausnaugh_2023}.
This pipeline evaluates a brightness of the star with difference imaging and fitting the point-spread function, and is expected to be more accurate than a simple aperture photometry for fainter sources ($\tess$ magnitude $\gtrsim$ 12).
FFIs record photo-electron count rate, and we converted it to energy flux density using the following relationship based on a documentation in an official web-page of $\tess$ Transients project\footnote{\url{https://tess.mit.edu/public/tesstransients}}:
\begin{equation}
    \frac{F}{I} = 1.61\times10^{-5}~\left[\frac{\mathrm{Jy}}{\mathrm{counts\,s^{-1}}}\right],
    \label{eq:cps2flux}
\end{equation}
where $I$ is the photo-electron count rate, and $F$ is an energy flux density in the $\tess$ pass-band.
The interval of FFIs was changed from 30 to 10 minutes at the end of the prime mission in July 2020 \citep{ricker_2021}.
We unified the sampling rate of the light curves by binning at 30-minutes widths during the analysis process (cf. \cref{sec:reduction}).
In addition, we obtained $\tess$ magnitude (Tmag) of target sources from $\tess$ Input Catalog \citep{stassun_2018,stassun_2019}.

\subsection{Long-term multi-wavelength light curves}\label{sec:lightcurves}
\begin{figure}
    \includegraphics[width=\columnwidth]{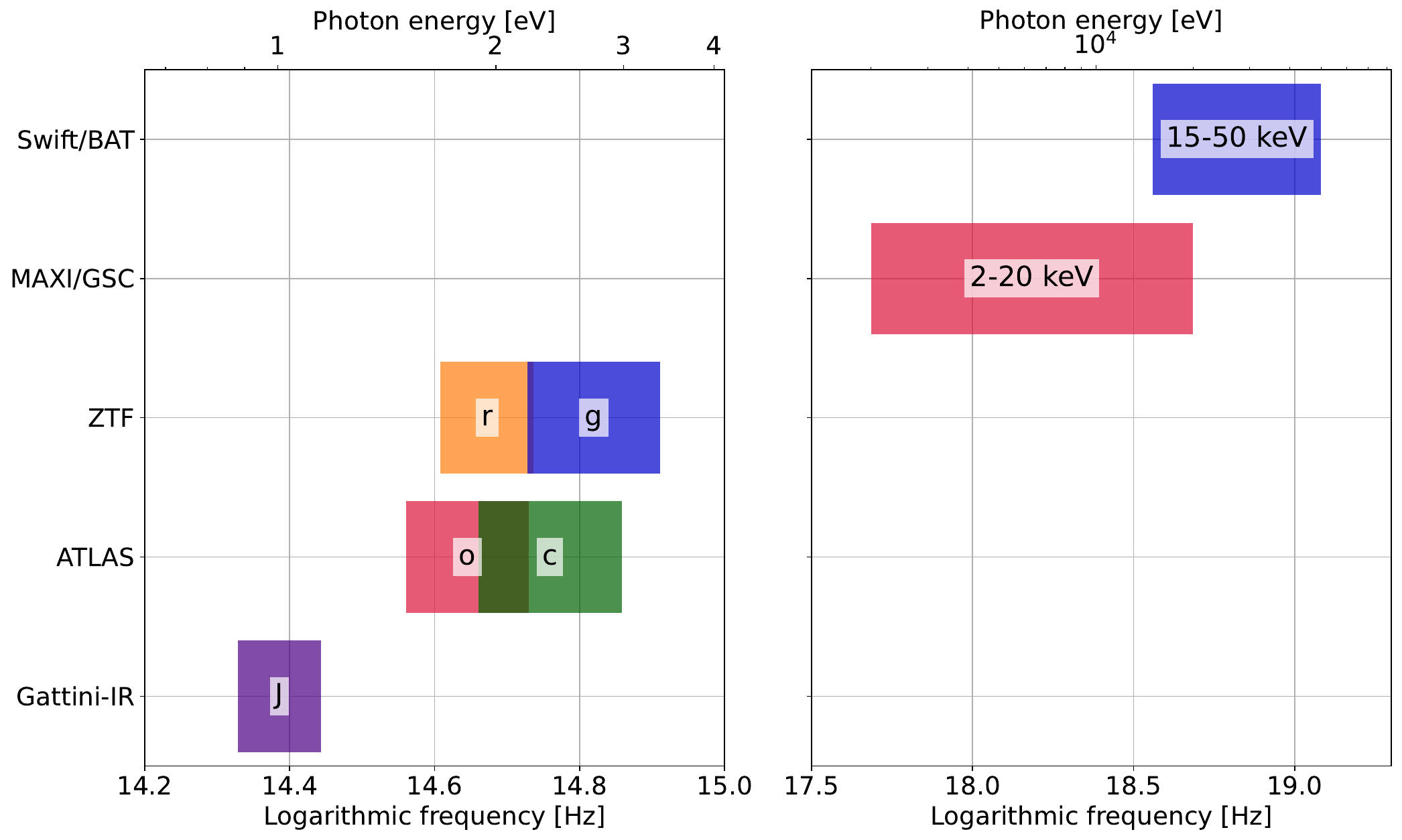}
    \caption{
        Energy bands of multi-wavelength light curves.
        Data of ZTF and ATLAS filters were obtained from the SVO Filter Profile Service (\url{http://svo2.cab.inta-csic.es/svo/theory/fps/index.php}).
        As for Gattini-IR, it is drawn with the value of the Cousins J-band filter.
    }
    \label{fig:passbands}
\end{figure}

We utilised X-ray light curves of $\swift$/BAT and MAXI/GSC, optical light curves of ZTF and ATLAS, and NIR light curves of Gattini-IR, covering MJD range of $\sim$ 58400--60000.
\cref{fig:passbands} shows energy bands of multi-wavelength light curves.

\subsubsection{Neil Gehrels Swift Observatory}
Neil Gehrels Swift Observatory ($\swift$; \citealp{gehrels_2004}) is a gamma-ray burst survey satellite operated by NASA.
We downloaded the Burst Alert Telescope (BAT) 15--50 keV daily light curves of target BeXBs, from $\swift$/BAT Hard X-ray Transient Monitor web-page \citep{krimm_2013}.
Seven days binning was performed to improve signal-to-noise ratio for each light curves.

\subsubsection{Monitor of All-sky X-ray Image}
Monitor of All-sky X-ray Image (MAXI; \citealp{matsuoka_2009}) is an X-ray camera installed onboard the International Space Station (ISS).
We obtained weekly-averaged Gas Slit Camera (GSC) 2--20 keV light curves covering 58000--60000 MJD for target BeXBs, via MAXI on-demand web interface \citep{nakahira_2012}. 

\subsubsection{Zwicky Transient Facility}
Zwicky Transient Facility (ZTF; \citealp{bellm_2019}) is an optical wide-field survey project using a CCD camera array attached to Samuel Oschin telescope at Palomar Observatory in California, operated by California Institute of Technology (Caltech).
We utilised $\gp$ and $\rp$-band light curves of the targets acquired from ZTF Public Data Release 15 (DR15; \citealp{masci_2019}), which covers the epoch of $\sim$ 58250--59900 MJD.
The light curve is not available only for RX J0440.9+4431 due to its brighter optical magnitude than the ZTF saturation magnitude ($\sim$ 12.5 AB).

\subsubsection{Asteroid Terrestrial-impact Last Alert System}
Asteroid Terrestrial-impact Last Alert System (ATLAS; \citealp{tonry_2018}) is a high-cadence optical all-sky survey project using four 50 cm telescope systems located in Haleakala, Mauna Loa (Hawaii), El Sauce (Chile), and Sutherland (South Africa).
We obtained $c$ and $o$-band light curves covering 58000--60000 MJD via ATLAS Forced Photometry web interface \citep{smith_2020}.
The saturation magnitude of ALTAS is $\sim$ 12.5 AB, about same as ZTF.
Thus, the light curve for RX J0440.9+4431, although it can be created, is not reliable.

\subsubsection{Gattini-IR}
Gattini-IR \citep{kishalay_2020} is a NIR wide-field survey project using 30 cm robotic $J$-band telescope system at Palomar Observatory, operated by Caltech.
Gattini-IR sweeps $\sim$ 7500 $\mathrm{deg}^2$ of the sky at one night with an upper-limit magnitude of $\sim$ 15.7 (AB), and scans the northern-sky observable from Palomar (Dec. $\gtrsim-30\degr$) with a cadence of $\sim$ 2 days.
The system is designed for the NIR-transient detection, and a real-time data processing pipeline (Gattini Data Processing System; GDPS) analyses observed images and detects the transients.
We obtained the $J$-band light curves of the targets generated by GDPS.
The data-set covers the epoch of $\sim$ 58400--59900 MJD.
\section{Data analysis}\label{sec:analysis}
We first estimated Be donor properties to evaluate the IR excess and the critical velocity.
Next, the IR excess was evaluated to investigate the evolution of the disks.
Finally, a flux periodicity was analysed applying the Lomb-Scargle technique \citep{lomb_1976,scargle_1982} to $\tess$ light curves.
We utilised \texttt{curve\_fit} function of \texttt{scipy.optimize} library \citep{virtanen_2020} for fitting, and \texttt{LombScargle} class of \texttt{astropy.timeseries} library \citep{astropy_2013, astropy_2018, astropy_2022} as an implementation of the Lomb-Scargle technique.

\subsection{Estimation of Be donor properties}\label{sec:be_property}
We estimated an effective temperature $\Teff$, radius $\Rstar$, and mass $\Mstar$ for each Be donor of target BeXBs by fitting of an optical spectral energy distribution (SED) with a simple blackbody model.
First, the optical SED of the target was created using catalogued optical magnitudes, a color excess $\ebv$, and a distance.
The reason for not using the NIR data (e.g. 2MASS) is that it is more susceptible to the IR excess than the optical band and the blackbody radiation may not be dominant.
We used the \citet{ccm89} extinction function to derive extinctions for each band.
This SED was then fitted with a model spectrum of a spherical blackbody to obtain $\Teff$ and $\Rstar$.
Finally, $\Mstar$ was calculated using the empirical relations of \citet{torres_2010} with some modifications:
\begin{align}
    X &= \log{\brkts{\frac{\Teff}{12.6~[\mathrm{kK}]}}},\label{eq:torres1}\\
    \log{\brkts{\frac{\Mstar}{\Msun}}} &= a_1 + a_2X + a_3X^2 + a_4X^3 + a_5\brkts{\log{g}}^2 + a_6\brkts{\log{g}}^3,\label{eq:torres2}\\
    \log{\brkts{\frac{\Rstar}{\Rsun}}} &= b_1 + b_2X + b_3X^2 + b_4X^3 + b_5\brkts{\log{g}}^2 + b_6\brkts{\log{g}}^3,\label{eq:torres3}
\end{align}
where $g$ is a surface gravity normalised in unit of $\mathrm{cm\,s^{-2}}$, $\Msun$ and $\Rsun$ are the solar mass and radius, respectively, and $a_{1\text{--}6}$ and $b_{1\text{--}6}$ are coefficients.
The base of the logarithm here is 10.
Values of $a_{1\text{--}6}$ and $b_{1\text{--}6}$ are listed in \cref{tab:coefficients}.
\begin{table}
    \caption{
        Coefficients of \cref{eq:torres2,eq:torres3} \citep{torres_2010}.
    }
    \begin{tabular}{ccc}\hline
        $i$ & $a_i$ & $b_i$\\\hline
        1 & $1.5689\pm0.058$ & $2.4427\pm0.038$\\
        2 & $1.3787\pm0.029$ & $0.6679\pm0.016$\\
        3 & $0.4243\pm0.029$ & $0.1771\pm0.027$\\
        4 & $1.139\pm0.24$ & $0.705\pm0.13$\\
        5 & $-0.1425\pm0.011$ & $-0.21415\pm0.0075$\\
        6 & $0.01969\pm0.0019$ & $0.02306\pm0.0013$\\\hline
    \end{tabular}
    \label{tab:coefficients}
\end{table}
Specifically, we solved \cref{eq:torres3} with a numerical technique to obtain $\log{g}$ from $\Teff$ and $\Rstar$, and calculated $\Mstar$ by substituting $\Teff$ and $\log{g}$ into \cref{eq:torres2}.
Note that we ignored metallicity [Fe/H] terms, which are in original equations, since all targets in this study are galactic sources.
We obtained the optical multi-band magnitudes from APASS DR10 \citep{henden_2018}, PS1 DR1 \citep{flewelling_2020}, or Table 2 of \cite{reig_2015}.
The indeterminacy of the result was estimated by the bootstrap re-sampling.
If multiple results were obtained by multiple catalogues, we selected the one for which $\Teff$ is most plausible to the typical value of its spectral type (cf. \cref{tab:spt}).

We estimated a blackbody $J$-band magnitude $\mbbj$, critical velocity $\vcrit$, and critical frequency $\nub$.
$\mbbj$ was obtained by extrapolating the obtained blackbody SED model to the $J$-band.
This is the expected $J$-band magnitude of the Be donor itself excluding the IR excess.
$\nub$ is a rotation frequency when spinning at $\vcrit$.
We calculated $\vcrit$ and $\nub$ assuming a sphere:
\begin{align}
    \vcrit =& \sqrt{\frac{G\Mstar}{\Rstar}},\\
    \nub =& \frac{\vcrit}{2\pi\Rstar},
\end{align}
where $G$ is the gravitational constant.
Since the rotation of the Be star does not reach $\vcrit$ in most cases \citep{rivinius_2013}, 
$\nub$ can be regarded as the upper limit of the rotational frequency.
If the contribution of the IR excess to the catalog optical magnitudes is not negligible, it causes the color to be redder and the magnitude to be larger, and results in underestimation of $\Teff$ and overestimation of $\Rstar$ and $\mbbj$.
Estimated properties are summarised in \cref{tab:stellar}.
At temperature $T\sim30000~\mathrm{K}$, the Planck distribution peak is at ultraviolet-band ($\nu\sim10^{15}~\mathrm{Hz}$), and $T$ change produces only a small difference in slope in optical band.
Therefore, the temperature estimation in this method is highly sensitive to small differences in the slope of the SED, resulting in a large error of $\Teff$ due to a slight uncertainty of $\ebv$.
Note that $\vcrit$ is consistent with the literature value of $\vsini$ (i.e. $\vcrit\gtrsim\vsini$) for all sources (cf. \cref{tab:target_list}).

\begin{table*}
    \caption{
        Estimated stellar properties of target BeXBs.
    }
    \begin{tabular}{ccccccccc}
        \hline
        Name & SpT & Src$^*$ & $\Teff$ & $\Rstar$ & $\Mstar$ & $\mbbj$ & $v_\mathrm{crit}$ & $\nub$\\
         & & & [kK] & [$\Rsun$] & [$\Msun$] & [AB] & [$\kmps$] & [$\dinv$]\\
        \hline
        4U0115+634 & B0.2Ve & APASS & $29\pm11$ & $9.0\pm2.1$ & $16\pm10$ & $12.9$ & $580\pm200$ & $1.3\pm0.6$\\
        GRO J2058+42 & O9.5-B0IV-Ve & PS1 & $38\pm15$ & $9.4\pm2.9$ & $29\pm23$ & $13.4$ & $760\pm320$ & $1.6\pm1.0$\\
        RX J0440.9+4431 & B0.2Ve & APASS & $28.6\pm7.0$ & $8.3\pm1.3$ & $15.6\pm6.6$ & $10.2$ & $600\pm130$ & $1.4\pm0.4$\\
        SAX J2103.5+4545 & B0Ve & [R\&F] & $32\pm12$ & $8.5\pm2.2$ & $20\pm14$ & $12.7$ & $660\pm260$ & $1.5\pm0.8$\\
        V0332+53 & O8.5Ve & APASS & $32.3\pm8.2$ & $10.6\pm2.2$ & $21\pm10$ & $12.6$ & $620\pm160$ & $1.1\pm0.5$\\
        \hline
        \multicolumn{8}{l}{$^*$ Sources of the optical magnitudes used to derive the properties.
        [R\&F] means \citet{reig_2015}.}
    \end{tabular}
    \label{tab:stellar}
\end{table*}

\subsection{IR excess evaluation}
\begin{figure}
    \includegraphics[width=\hsize]{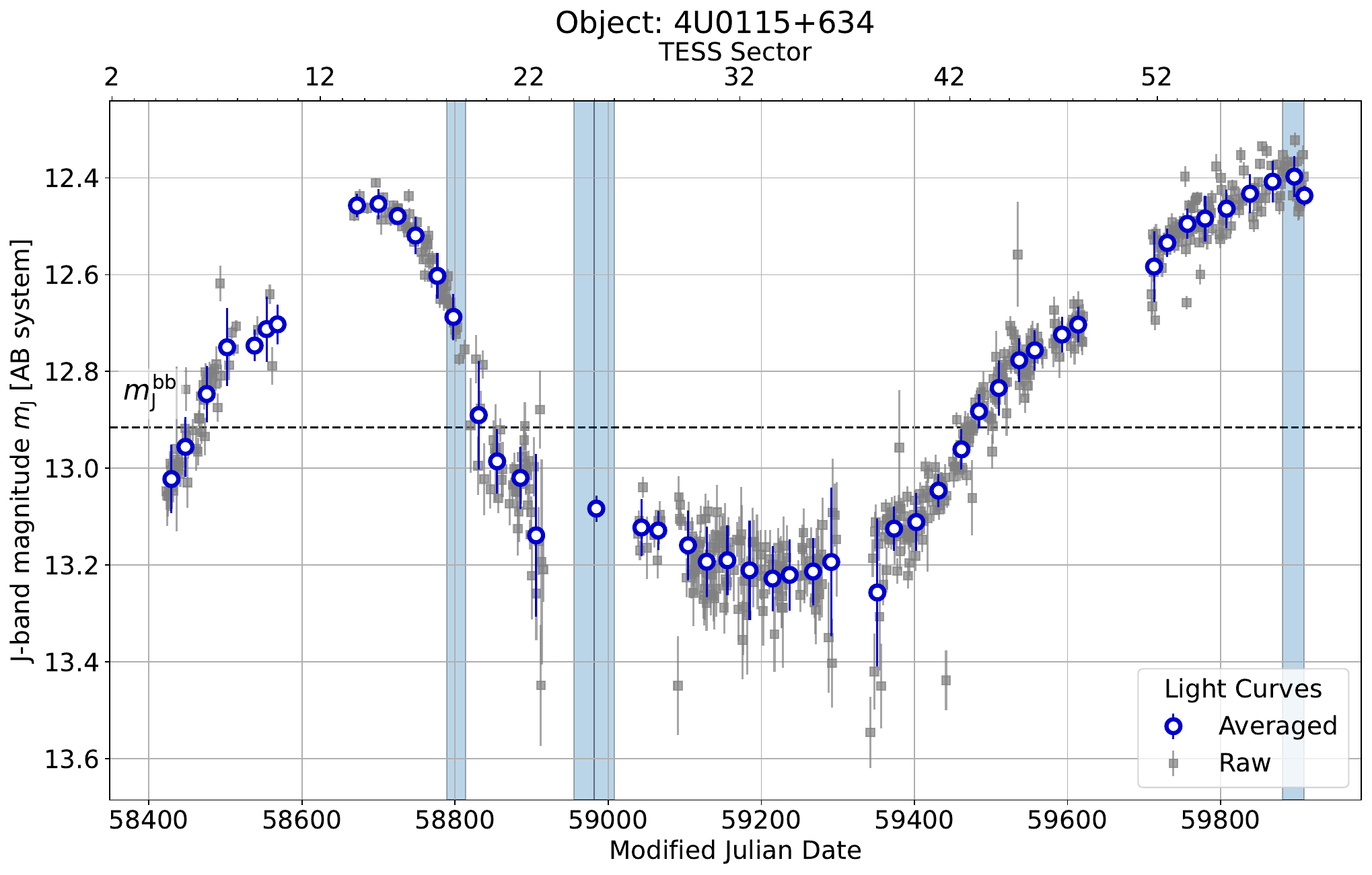}
    \caption{
        IR excess evaluation of 4U0115+634.
        Gray square and blue circular markers correspond to the raw Gattini-IR $J$-band light curve and the sector-by-sector averaged light curve, respectively.
        The horizontal dashed line is $\mbbj$ which is the estimated IR excess excluded magnitude, and the blue shaded regions represent sectors where $\tess$ observed this source.
        We used the difference between the blue points and $\mbbj$ line as a measure of the IR excess.
        The fact that the light curve is sometimes systematically below $\mbbj$ (e.g. 58800--59500 MJD) may be due to an overestimation of $\mbbj$.
    }
    \label{fig:irexcess}
\end{figure}

We evaluated the IR excess as a measure of disk development.
The $J$-band flux observed by Gattini-IR was averaged per sector and converted into magnitudes, defined as $\mj$.
We used $\mj-\mbbj$ to evaluate the amount of the IR excess (\cref{fig:irexcess}).
The smaller the value of $\mj-\mbbj$, the greater the IR excess.
The Gattini-IR magnitudes were systematically dimmer than $\mbbj$ in 4U0115+634.
This can be qualitatively explained by overestimation of $\mbbj$ due to some factors such as underestimation of $\Teff$ (cf. \cref{sec:be_property}).
In any case, the IR excess evaluated in this way can be used to discuss long-term variations of one source, but not for cross-source comparisons.

\subsection{Analysis of flux periodicity}\label{sec:lombscargle}
We made and analyzed amplitude spectra of $\tess$ light curves for each sector of each targets to investigate the flux periodicity.
This analysis was carried out with the following procedure:
\begin{enumerate}
    \item Data reduction to raw $\tess$ light curves,
    \item Amplitude spectrum estimation using the Lomb-Scargle technique,
    \item Decomposition of the spectrum into peaks and the red noise.
\end{enumerate}

\subsubsection{Data reduction}\label{sec:reduction}
\begin{figure}
    \includegraphics[width=\hsize]{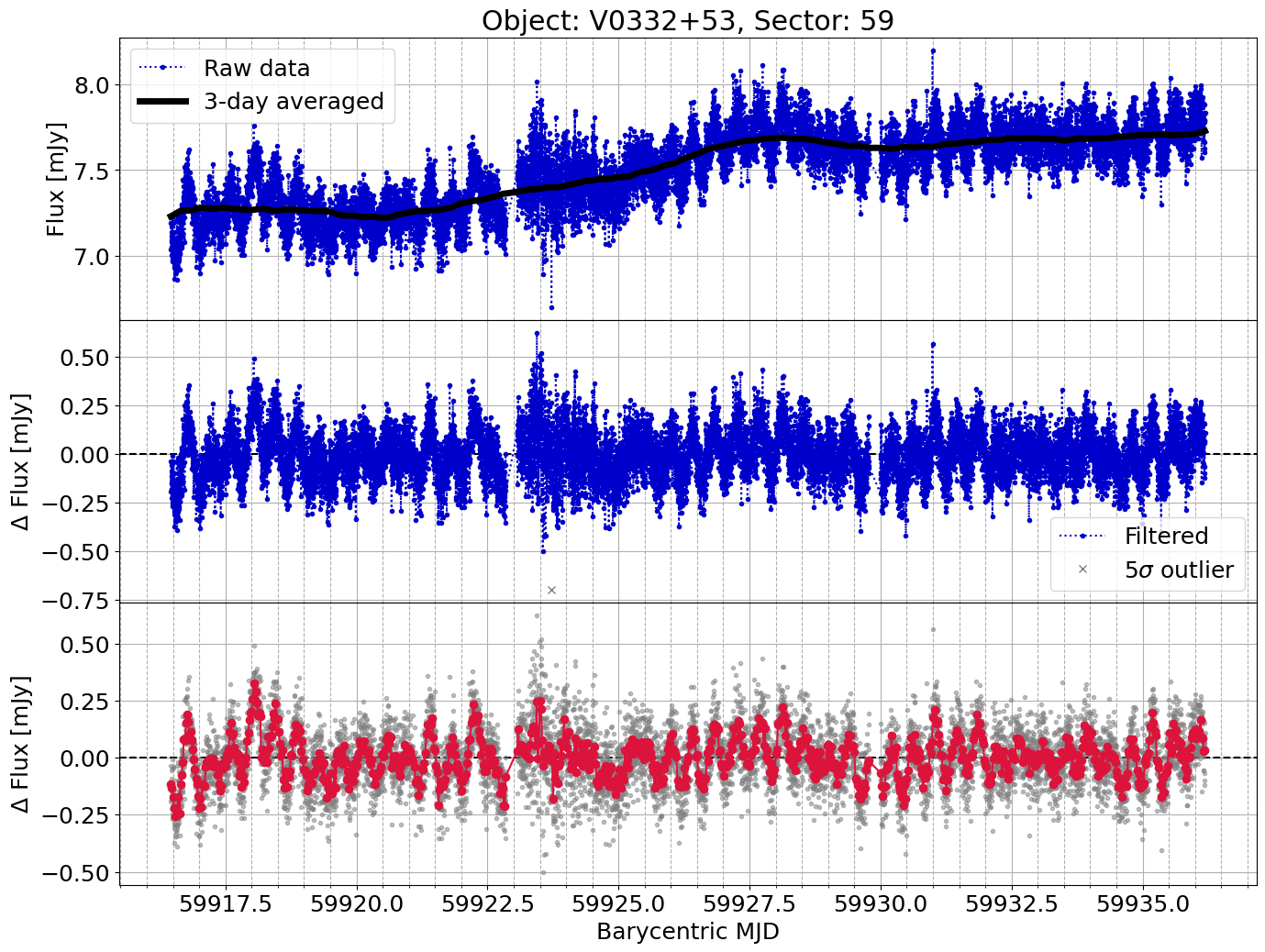}
    \caption{
        Data reduction to the $\tess$ light curve of V0332+53, sector 59.
        The three panels correspond to the 3-day detrending, 5-sigma clipping, and 30-min binning processes, respectively.
    }
    \label{fig:reduction}
\end{figure}

The data reduction process consists of three stages: detrending, 5-sigma clipping, and 30-min binning.
First, we subtracted the moving averaged flux from the light curve to remove the long-term flux variation in time scale of $\gtrsim$ several days.
For convenience, we call this process as `detrending'.
The width of the detrending window was set to 3 days except for SAX J2103.5+4545.
For SAX J2103.5+4545, the width was set to 1 day because the 3-day detrending was insufficient to remove long-term variability.
Next, 5-sigma clipping was performed to remove outliers.
Finally, we binned light curves in 1/48 days ($\sim$ 30 min) width to minimise the effect of differences in the sampling rate of FFIs between the prime and extended missions \citep{ricker_2021}.
\cref{fig:reduction} shows the data reduction process to the $\tess$ light curve of V0332+53, sector 59.

\subsubsection{Amplitude spectrum}\label{sec:fourier}
\begin{figure}
    \includegraphics[width=\hsize]{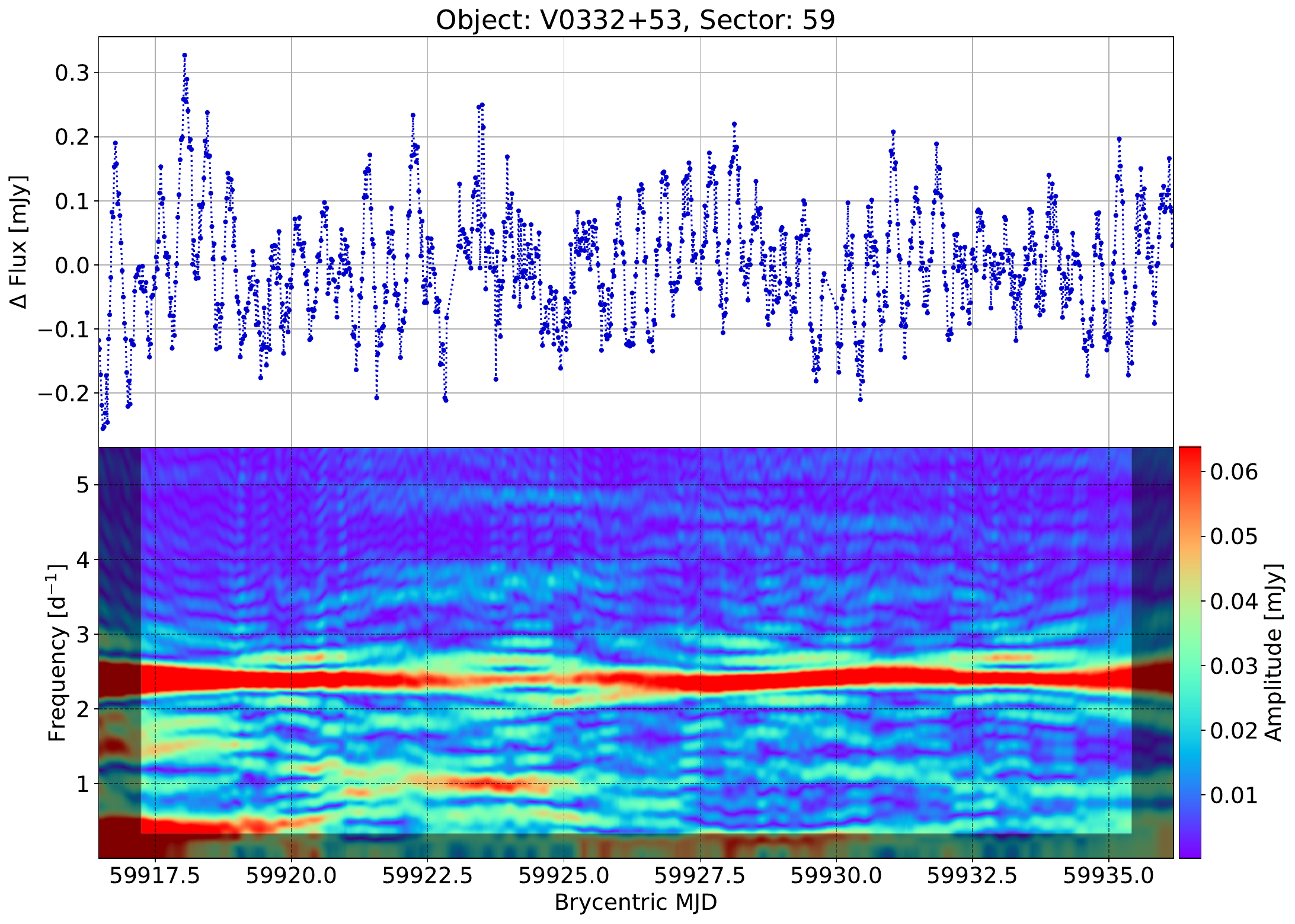}
    \caption{
        Dynamic amplitude spectrum of V0332+53, sector 59.
        The top panel is a light curve, and the bottom is a dynamic spectrum.
        The shaded area is where the frequency is less than 1/3 $\dinv$ or the number of extracted light curve points is less than 160.
    }
    \label{fig:dynamic}
\end{figure}

\begin{figure}
    \includegraphics[width=\hsize]{
        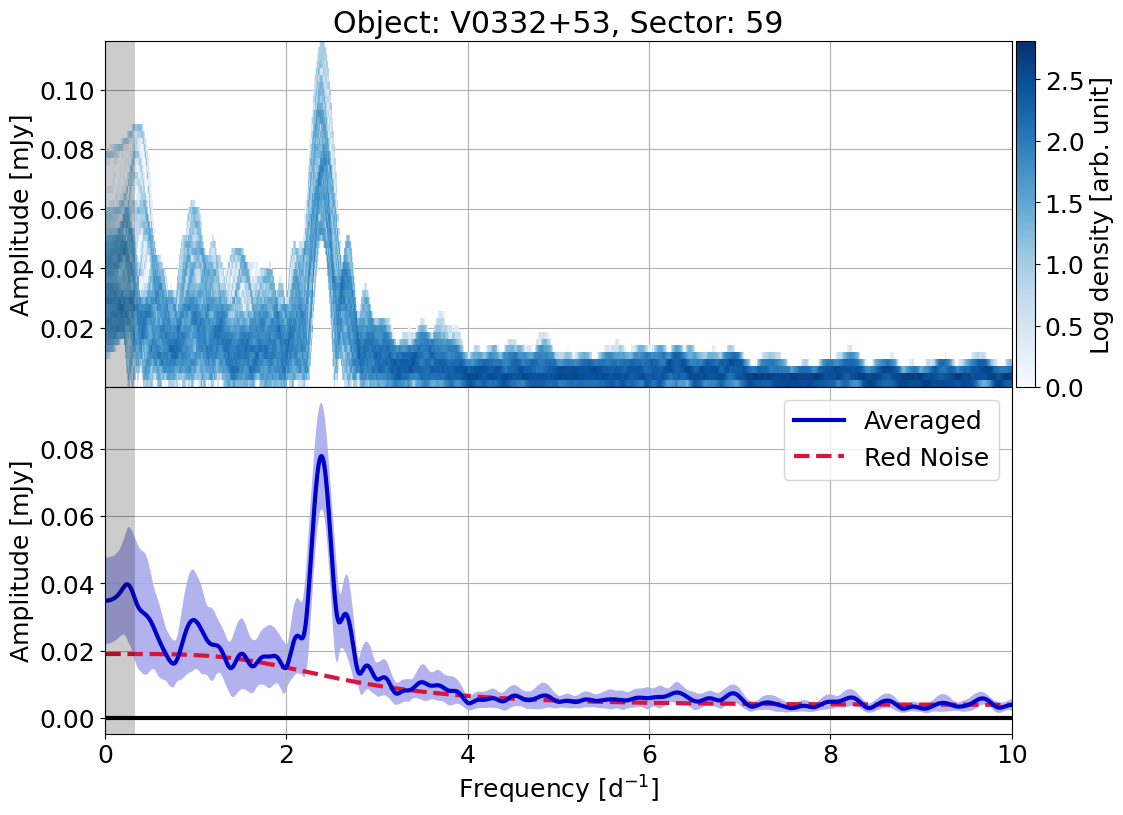
    }
    \caption{
        Averaged amplitude spectrum of V0332+53, sector 59.
        The top panel shows a 2D histogram of dynamic amplitude spectrum (\cref{fig:dynamic}) projected along the time axis, and the bottom panel shows an averaged amplitude spectrum.
        The averaged amplitude spectrum can be paraphrased as the centroid of the histogram for each frequency.
        `Log density' is the logarithm of density (arbitrary units) and its base is 10.
        The left shaded area is where the frequency is less than 1/3 $\dinv$.
        See also \cref{sec:decomposition} for the description of the red noise.
    }
    \label{fig:average}
\end{figure}

We applied the Lomb-Scargle technique to the reduced light curves.
Note that we did not use a power spectrum (periodogram), but an amplitude spectrum calculated by the following transformation (cf. \citealp{aerts_2021}):
\begin{equation}
    A(\nu) = \sqrt{\frac{4}{N}P(\nu)},
\end{equation}
where $\nu$ is a frequency, $A(\nu)$ is the amplitude spectrum, $N$ is a number of points in the light curve, and $P(\nu)$ is the power spectrum.
We first made dynamic amplitude spectra.
A dynamic amplitude spectrum is obtained by taking a portion of the light curve with an appropriate window and obtaining its spectrum by applying the Lomb-Scargle while moving the window.
That shows the variation of the spectrum.
The window width was set to 5 days and moved by 1/48 days.
Note that oscillations with a period equal to or greater than the width of the detrending window had removed in the data reduction process.
\cref{fig:dynamic} is a dynamic amplitude spectrum of V0332+53, sector 59.
If the number of points in the extracted light curve was less than 160, the obtained spectrum was considered to be unreliable, where the number 160 is 2/3 of the number of points expected from the sampling rate (48 $\dinv$) and the window width (5 days).
At the same time, we evaluated the contribution of photometric errors to the spectrum by creating mock light curves which fluctuate randomly with a Gaussian-distribution over the width of the photometric error and calculating their spectrum.
Note that the photometry pipeline we used has a tendency to underestimate errors by about 10 per cent and up to 50 per cent, compared with estimated from the light curve fluctuations.
Finally, we averaged each dynamic spectrum along the time-axis direcion (\cref{fig:average}).
In this calculation, spectrum with points fewer than 160 on the light curve were excluded.
We defined this as "averaged amplitude spectrum", and considered it to be the representative amplitude spectrum of the object in the sector.
The reason why Lomb-Scargle is not simply applied to the entire light curve for each sector is to reduce the effect of several days of missing points on some light curves.

\subsubsection{Decomposition into peaks and red noise}\label{sec:decomposition}
\begin{figure}
    \includegraphics[width=\hsize]{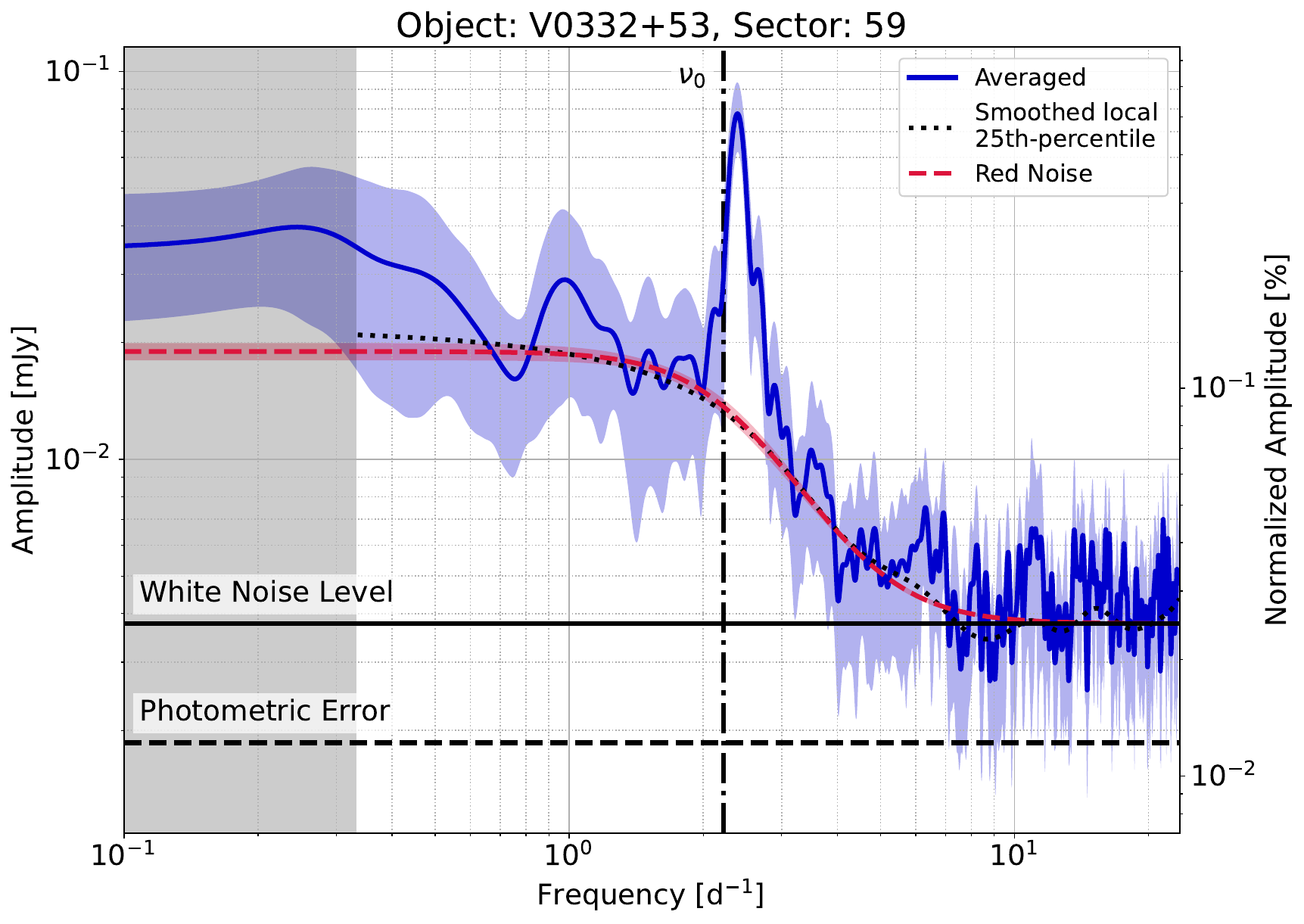}
    \caption{
        Red noise estimation of V0332+53, sector 59.
        The smoothed local 25th percentile is the roughly estimated continuum, and white noise level is the constant component of the red noise model.
        `Photometric Error' means a spectrum estimated from the photometric error.
        `Normalized Amplitude' is the ratio of amplitude to the flux converted from Tmag.
        The left shaded area is where the frequency is less than 1/3 $\dinv$.
    }
    \label{fig:rednoise}
\end{figure}

To extract the peaks from the spectrum, we estimated a red-noise component of the spectrum.
First, we made a local 25th percentile (first quartile) curve by extracting a portion of the spectrum with a window moving across all frequencies, and calculating the 25th percentile for each window.
Then the 25th percentile curve is smoothed with a Gaussian filter.
The obtained curve can be considered to be a rough estimate of the continuum.
The reason for using the 25th percentile rather than the mean or median is the nature of the amplitude spectrum.
Unlike random errors, which can be both positive and negative, the peaks in the amplitude spectrum are almost always positive.
Therefore, the mean or median are expected to be systematically larger than the continuum.
Finally, the red noise spectrum was obtained by model fitting to the roughly estimated continuum.
We utilised the model used in \citet{bowman_2019} and \citet{naze_2020} with some modifications.
The model consists of a Lorentzian-like term and a constant term:
\begin{equation}
    A_\mathrm{RN}(\nu) = \sqrt{\frac{a_0^2}{1+\left(\nu/\nu_0\right)^\gamma} + C^2},\label{eq:rednoise}
\end{equation}
where $\nu$ is a frequency, $A_\mathrm{RN}$ is a red noise amplitude spectrum, $a_0$ is a scaling factor, $\nu_0$ is a characteristic frequency, $\gamma$ is an exponent, and $C$ is a white noise level.
When $\gamma=2$, the first term is the Lorentzian.
If both sides of \cref{eq:rednoise} are squared, the equation agrees with \citet{bowman_2019} and \citet{naze_2020}.
The square root is applied to make the dimension of both sides amplitude rather than power.
\cref{fig:rednoise} shows the red noise estimation of V0332+53, sector 59.
Incidentally, the error-derived amplitude (cf. \cref{sec:fourier}) was smaller than the white noise level for all light curves as shown in \cref{fig:rednoise}, even if putting into account the underestimation of the error.

The peaks were extracted by subtracting the red noise from the amplitude spectrum.
We determined that the peaks whose amplitude were larger than twice the red noise were significant.
For each source, we evaluated the amplitudes for all sectors at frequencies for which a significant peak was detected in at least one of the sectors.
The amplitude of the detected peak was normalised by the flux calculated from the Tmag of the source.
This normalisation is based on the assumption that Tmag is the typical magnitude of the source in the $\tess$ pass-band, and ignores long-term optical variations.

\section{Results}\label{sec:result}

\subsection{Long-term multi-wavelengths activities}\label{sec:lc_result}
\begin{figure}
    \includegraphics[width=\columnwidth]{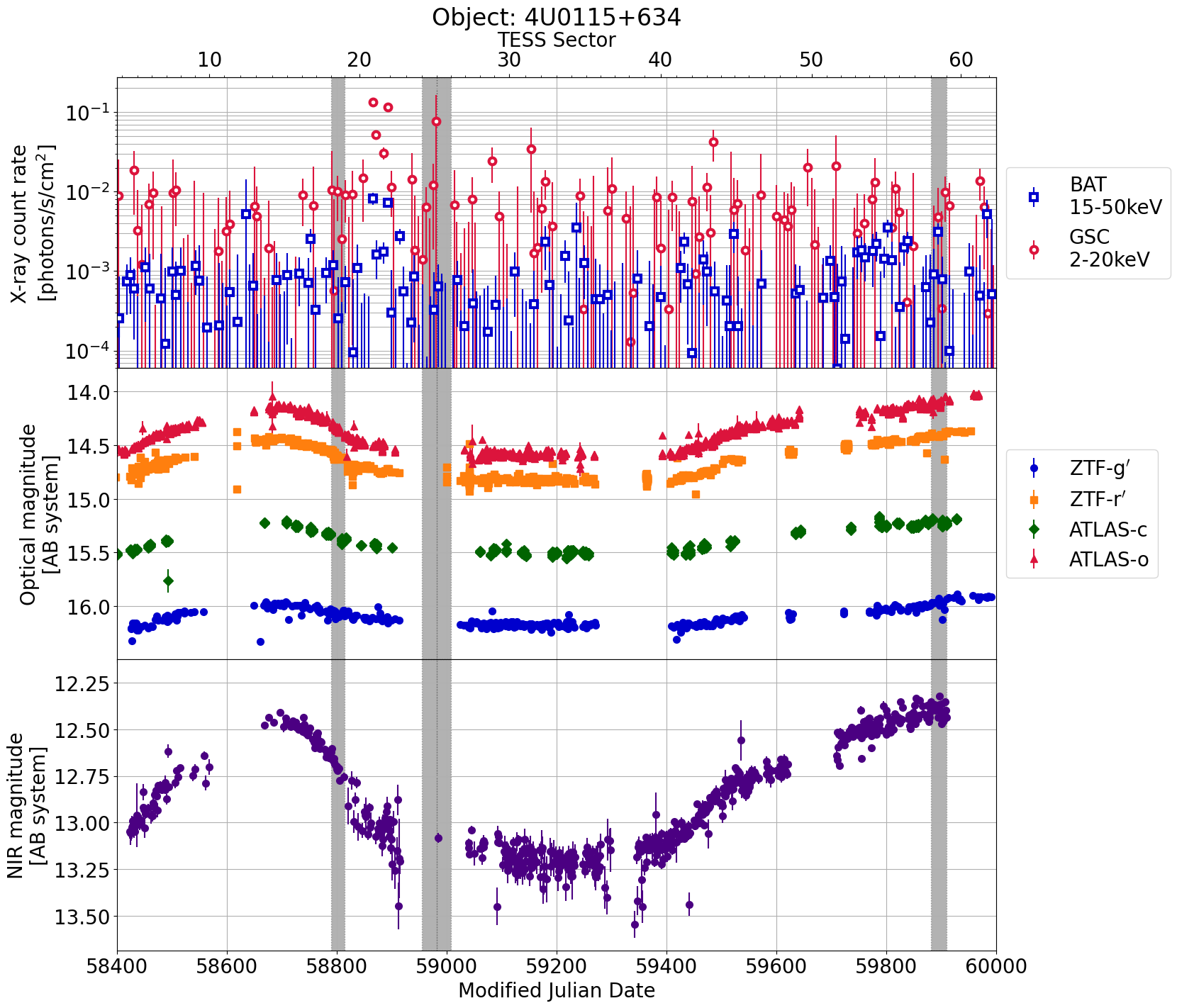}
    \caption{
        Multi-wavelength light curves of 4U0115+634.
        The gray shaded regions correspond to $\tess$ observed sectors.
        An X-ray outburst can be confirmed at $\sim$ 58900 MJD, which is consistent with the Atel report \citep{nakajima_2020}.
    }
    \label{fig:lc_4U0115}
\end{figure}

\begin{figure}
    \includegraphics[width=\columnwidth]{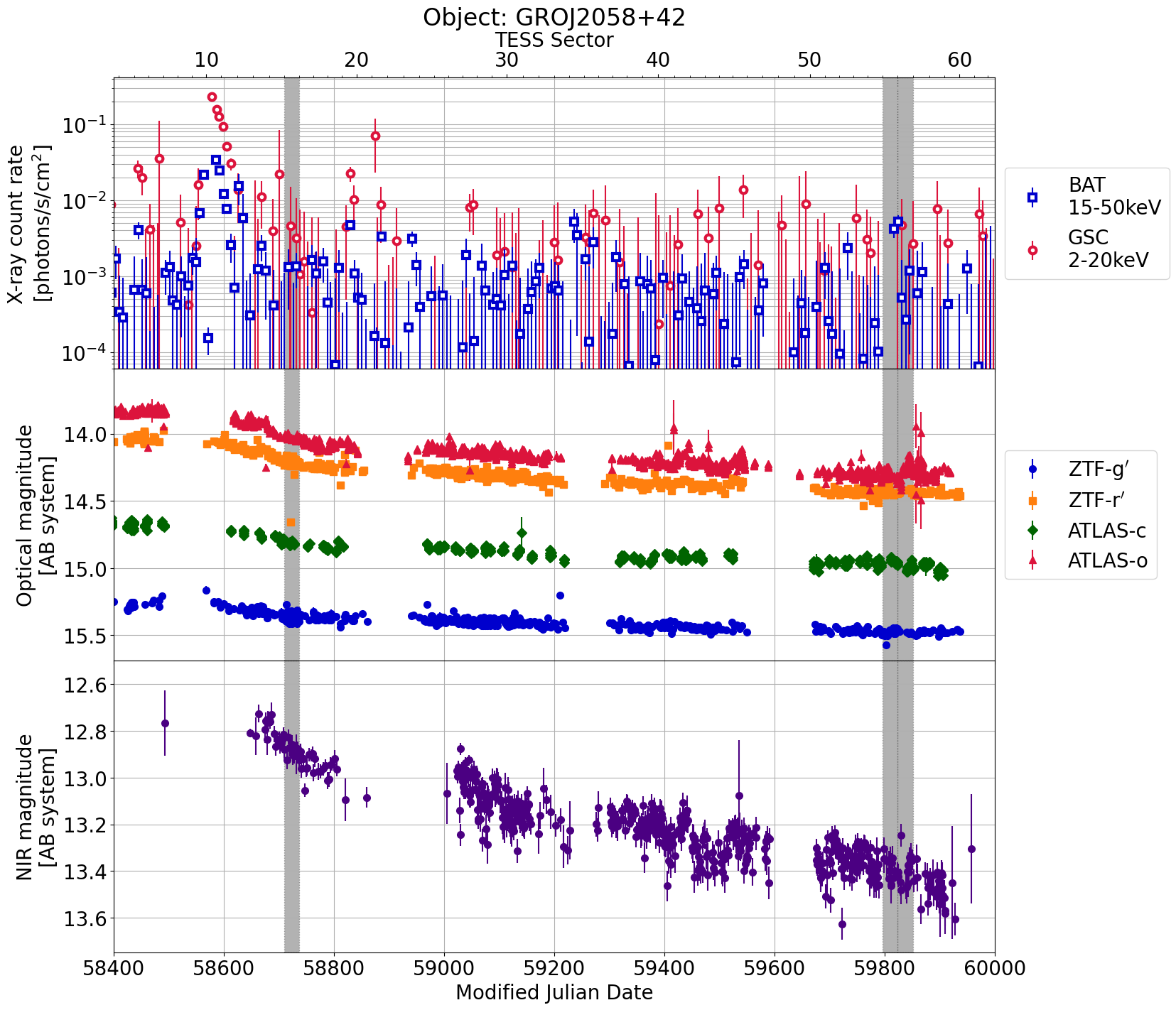}
    \caption{
        A similar figure to \cref{fig:lc_4U0115} for GRO J2058+42.
        An X-ray outburst can be confirmed at $\sim$ 58500 MJD, which is consistent with the GCNC report \citep{barthelmy_2019}.
    }
    \label{fig:lc_GROJ2058}
\end{figure}

\begin{figure}
    \includegraphics[width=\columnwidth]{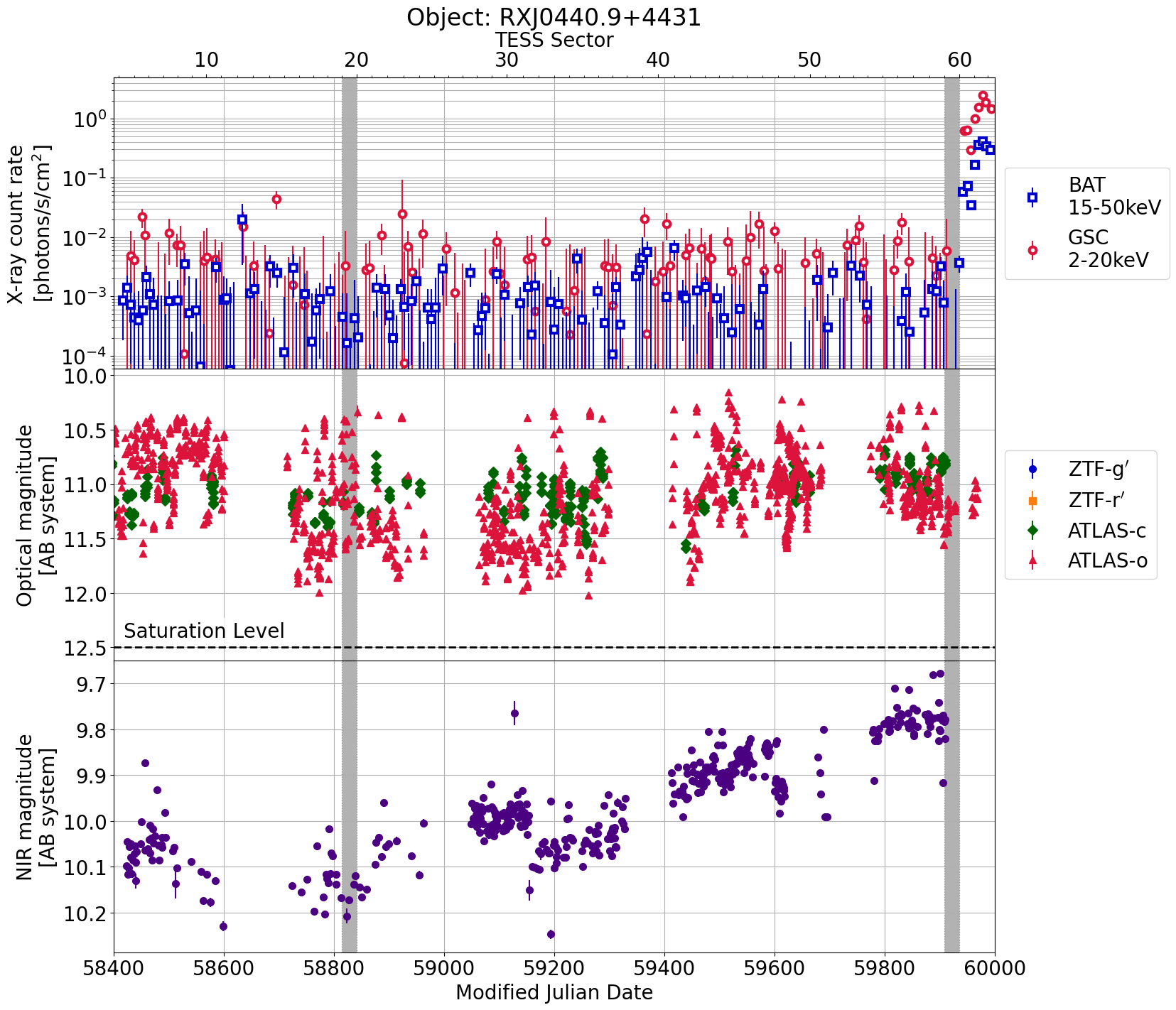}
    \caption{
        A similar figure to \cref{fig:lc_4U0115} for RX J0440.9+4431.
        Optical light curves exceeded the saturation level and are unavailable for analysis.
        An X-ray outburst can be confirmed at $\sim$ 60000 MJD, which is consistent with the Atel report \citep{nakajima_2022}.
    }
    \label{fig:lc_RXJ0440}
\end{figure}

\begin{figure}
    \includegraphics[width=\columnwidth]{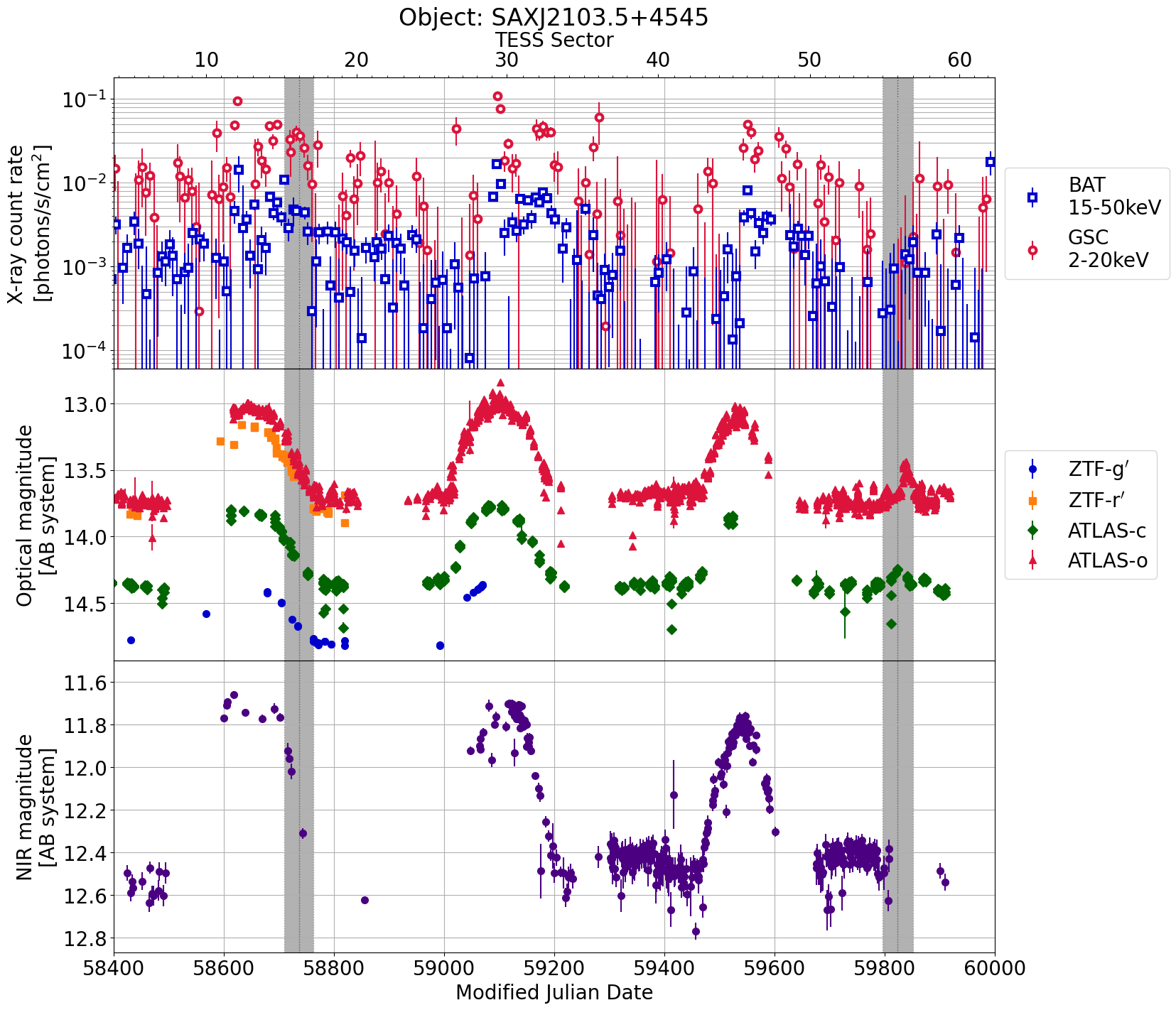}
    \caption{
        A similar figure to \cref{fig:lc_4U0115} for SAX J2103.5+4545.
        Recurrent X-ray/OIR outbursts can be confirmed at $\sim$ 58600, 59000, and 59500 MJD.
    }
    \label{fig:lc_SAXJ2103}
\end{figure}

\begin{figure}
    \includegraphics[width=\columnwidth]{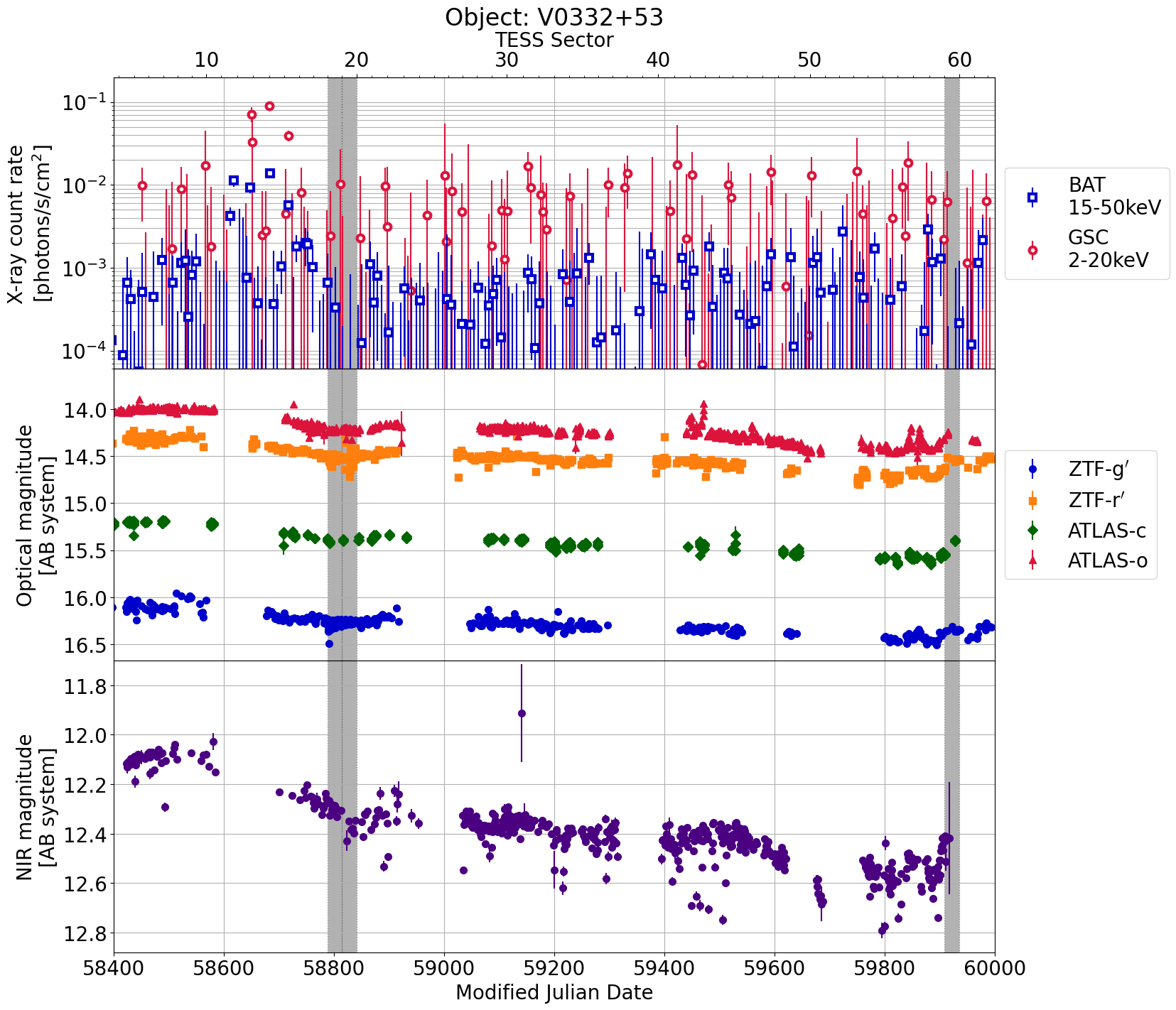}
    \caption{
        A similar figure to \cref{fig:lc_4U0115} for V0332+53.
        An X-ray outburst can be confirmed at $\sim$ 58700 MJD.
    }
    \label{fig:lc_V0332+53}
\end{figure}

\cref{fig:lc_4U0115,fig:lc_GROJ2058,fig:lc_RXJ0440,fig:lc_SAXJ2103,fig:lc_V0332+53} are long-term multi-wavelength light curves of five sources.
They exhibited significant X-ray outbursts in both 2--20 keV and 15--50 keV light curves, and OIR variability of $\gtrsim$ 0.5 mag.
The X-ray light curves of SAX J2103.5+4545 during outbursts show two components: a sharp peak with a duration of about 10 days and a relatively gentle bump of about 100 days.
This feature was also seen in previous outbursts of this source (cf. \citealp{camero_2014}).

The Crab pulsar has a luminosity of $\sim10^{37}~\eps$ \citep{kirsch_2005}, a distance of $\sim2~\mathrm{kpc}$ \citep{lin_2023}, and is detected by the GSC on the order of $1~\cps$ \citep{morii_2011}.
By comparing GSC count rates and distances of five targets with those of the Crab pulsar, the peak luminosities of these outbursts can be roughly estimated to be $\gtrsim10^{37}~\eps$.
In addition, the orbital periodicity expected for normal outbursts was not confirmed, and the duration of outbursts was $\sim$ 100 days, which is longer than orbital periods other than RX J0440.9+4431.
Outbursts of SAX J2103.5+4545 appear to be periodic, but their interval ($\sim$ 200 days) is not consistent with its orbital period (12.67 days).
Therefore, all detected X-ray outbursts can be classified as giant outbursts.

\begin{figure}
    \includegraphics[width=\columnwidth]{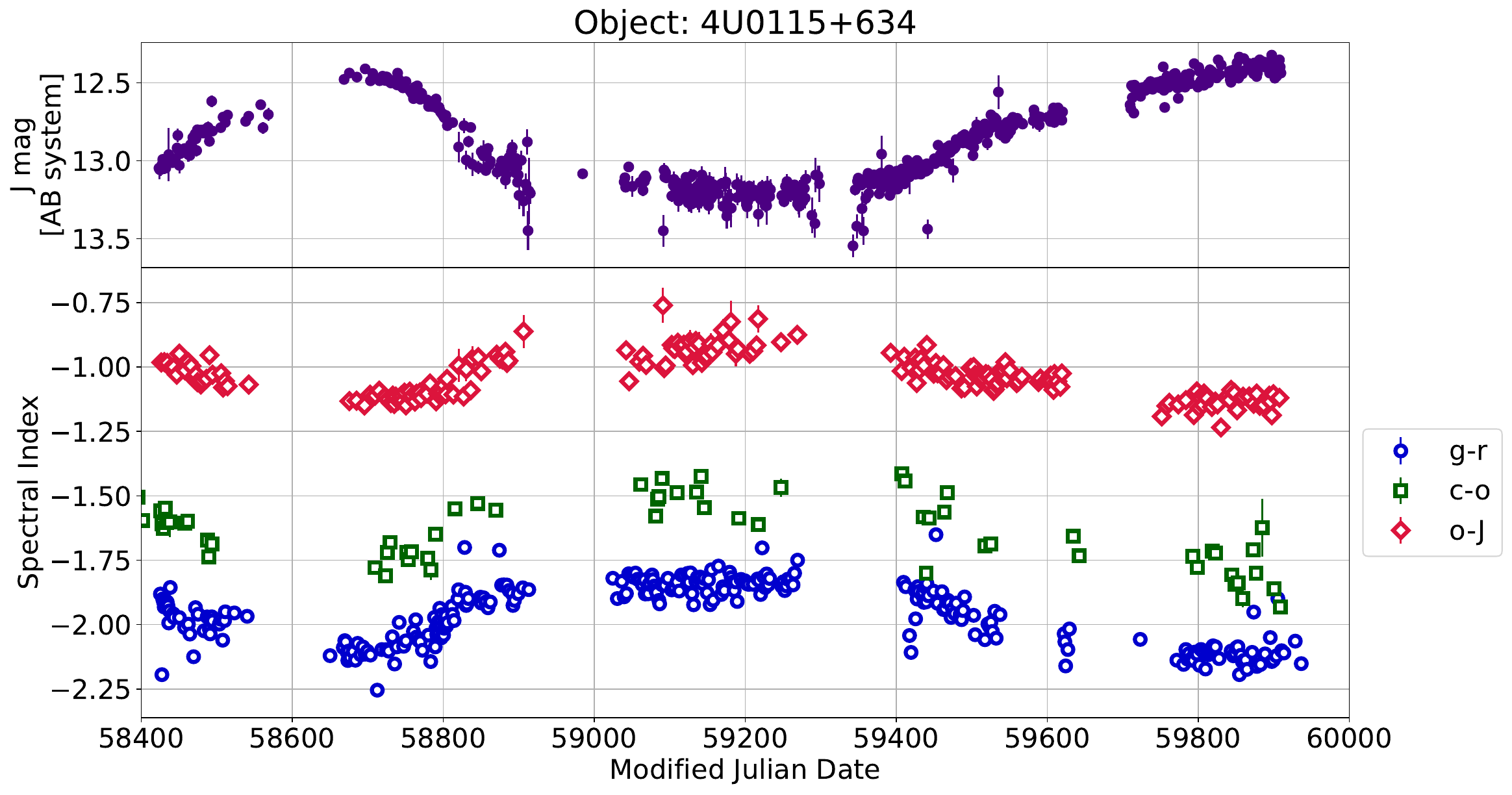}
    \caption{
        OIR color variations of 4U0115+634.
        Top and bottom panel show the J-band light curve and OIR color variations, respectively.
        The vertical axis in the bottom panel is the spectral index, which is redder at the bottom and bluer at the top.
    }
    \label{fig:color_4U0115}
\end{figure}

The time scales for all OIR light curve variations were about hundreds of days, which are consistent with the typical time scale of the circumstellar disk evolution.
\cref{fig:color_4U0115} shows the OIR color variations of 4U0115+634.
The spectral index shown in the figure represents the color, and is calculated for given two bands as follows:
\begin{equation}
    \alpha\brkts{\nu_0,\nu_1} = \frac{\log{\brkts{f_{\nu_0}/f_{\nu_1}}}}{\log{\brkts{\nu_0/\nu_1}}},
\end{equation}
where $\nu_0$ and $\nu_1$, and $f_{\nu_0}$ and $f_{\nu_1}$ are photon frequencies and fluxes in the given two bands, respectively, and $\alpha$ is a spectral index.
This corresponds to the slope of the line connecting two points in the double logarithmic SED, and a larger value indicates a bluer spectrum.
The figure shows a "redder when brighter" tendency that are common for OIR variations of Be stars.
This tendency was confirmed for all targets except RX J0440.9+4431 whose optical light curves were unavailable.
Therefore, all of the observed long-term OIR variations can be attributed to the disk evolution.
Thus, observed $J$-band flux, and hence $\mj-\mbbj$, can be used as a measure of the disk development.

Recent studies on giant outbursts \citep{martin_2011,martin_2014,okazaki_2013} suggested that such events should occur when a neutron star passes near a sufficiently developed disk that extends to the orbit of the neutron star.
Since the orbital period of BeXBs is typically shorter than the time scale of the disk growth, giant outbursts are expected to occur at or shortly before the maximum IR excess.
However, all detected giant outbursts were initiated months to hundreds of days after the peak of the $J$-band light curve.
The same was observed in the 2017 giant outburst of Swift J0243.6+6124 \citep{alfonso_2024}.
This discrepancy can be explained by assuming that the peak IR excess and the maximum outer disk radius are not simultaneous.
Since the mass is supplied to the disk from the Be star, i.e., from the innermost edge of the disk, there should be a time lag of about the viscous time scale between the mass supply and the outward extension of the disk.
In other words, the total mass and outermost radius never reach the maximum at the same time, and the former does earlier.
In addition, since the volume emissivity of thermal bremsstrahlung is proportional to the square of the electron density, IR excess should depend on both disk radius and mass, and the peak of IR excess should also be before the radius peak.
Hence, the observed delay can be attributed to the time lag between mass supply and disk extension.

\subsection{Amplitude spectra}
\begin{figure}
    \includegraphics[width=\columnwidth]{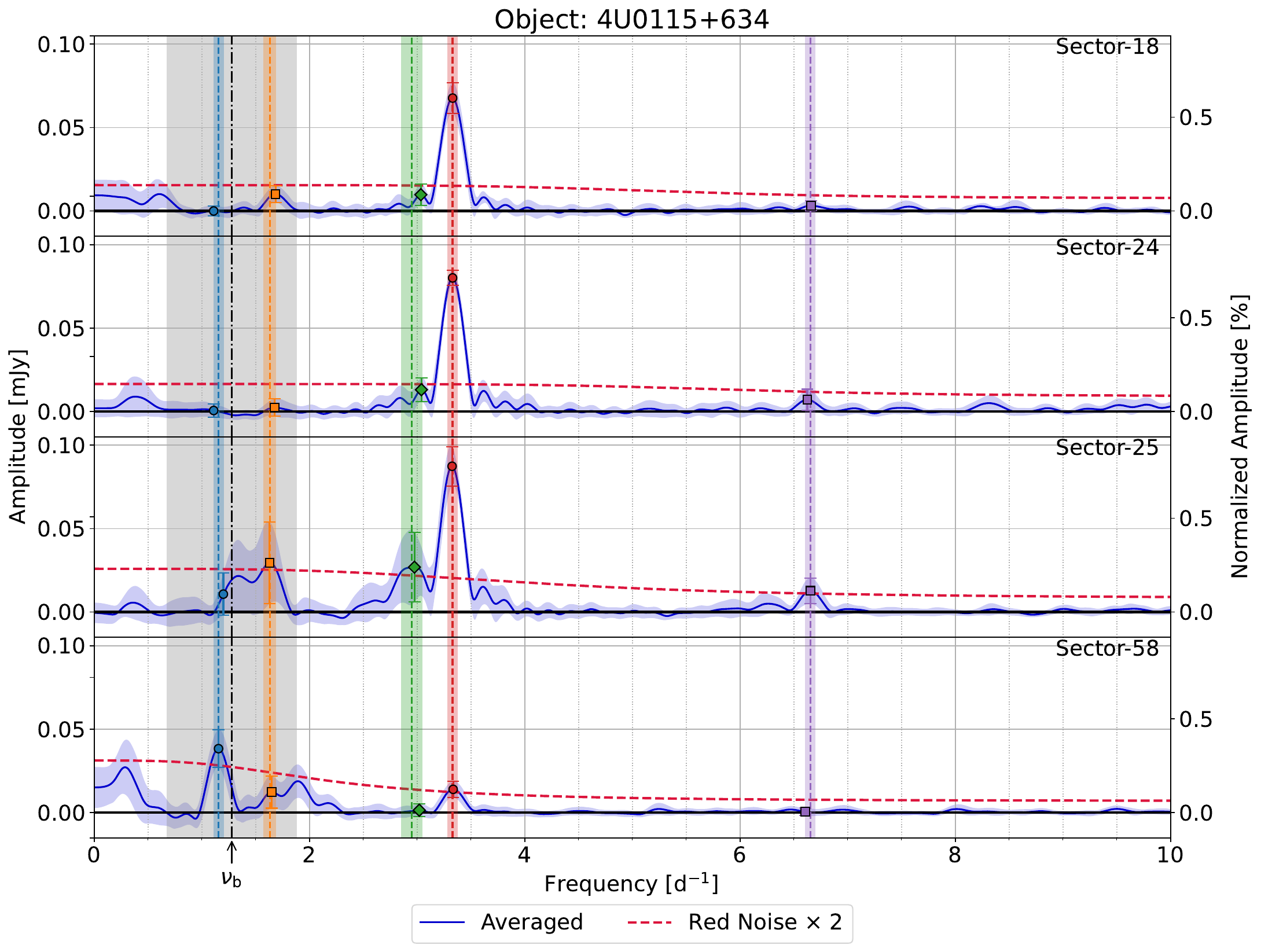}
    \caption{
        Red noise subtracted amplitude spectra of 4U0115+634.
        `Normalized Amplitude' is the amplitude normalised by the flux calculated from the Tmag of the source.
        `Red Noise $\times$ 2' line is the threshold of the peak detection.
        Vertical dashed lines indicate positions of detected peaks.
    }
    \label{fig:spectra_4U0115}
\end{figure}

\begin{figure}
    \includegraphics[width=\columnwidth]{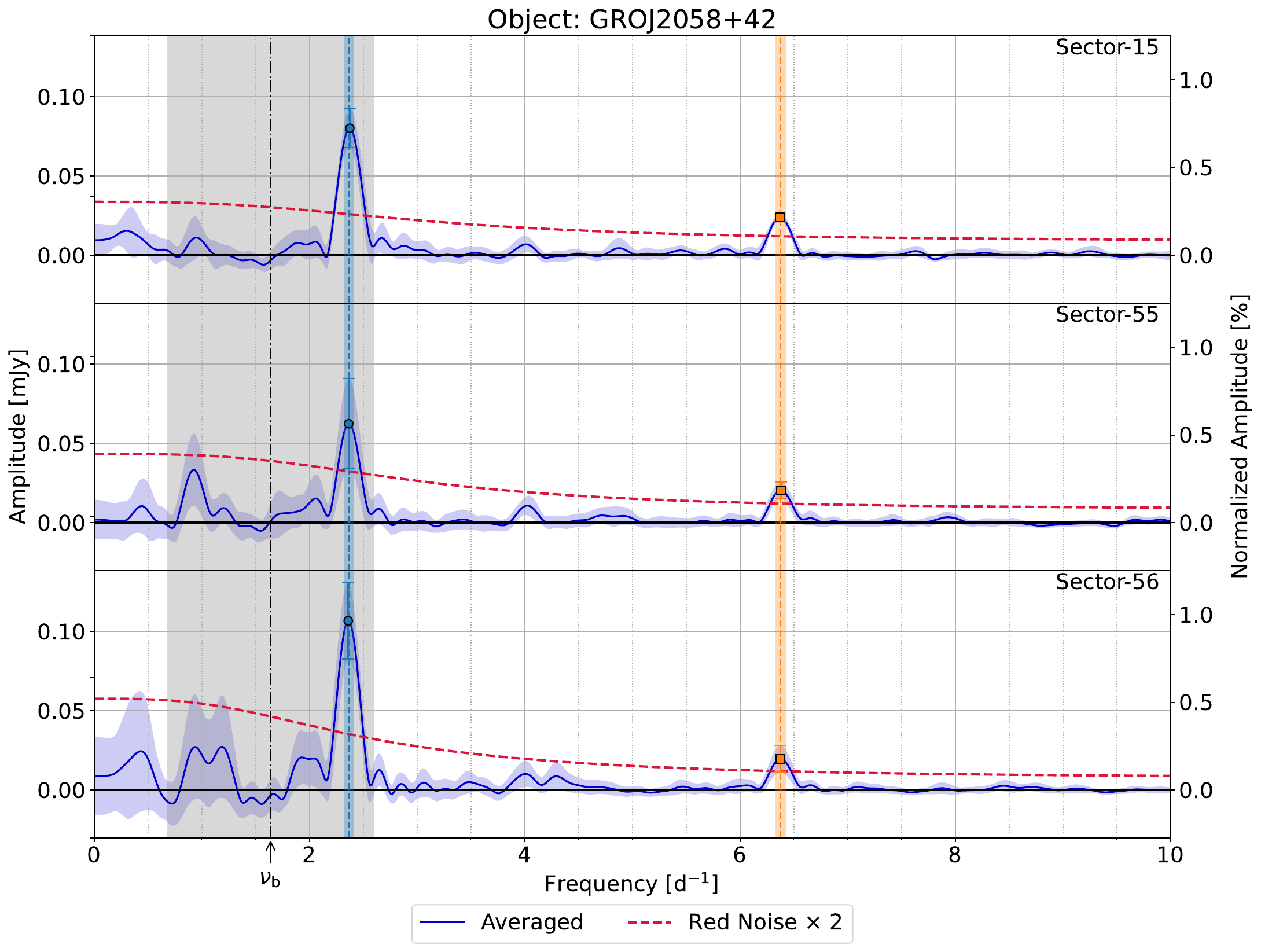}
    \caption{
        A similar figure to \cref{fig:spectra_4U0115} for GRO J2058+42.
    }
    \label{fig:spectra_GROJ2058}
\end{figure}

\begin{figure}
    \includegraphics[width=\columnwidth]{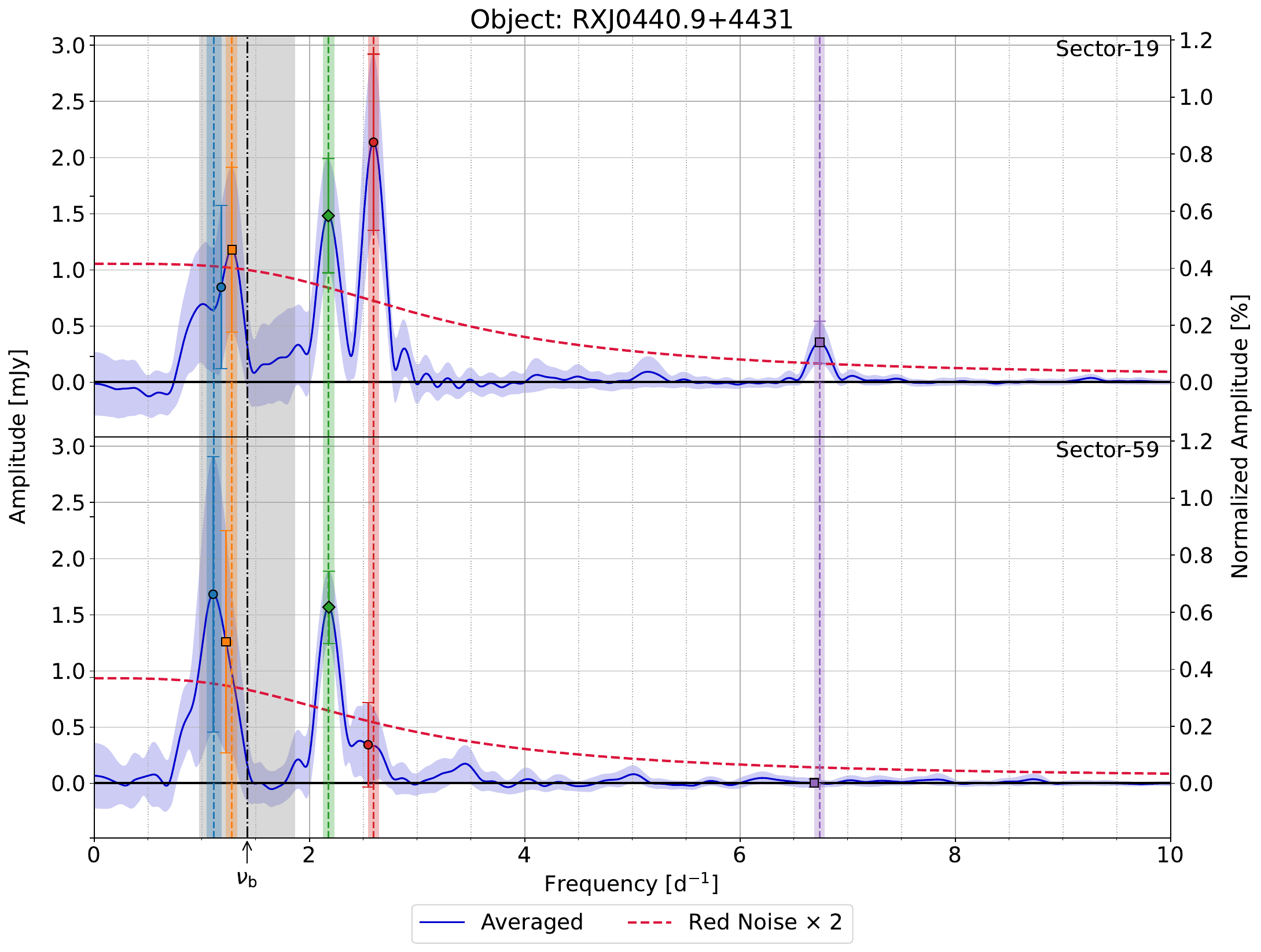}
    \caption{
        A similar figure to \cref{fig:spectra_4U0115} for RX J0440.9+4431.
    }
    \label{fig:spectra_RXJ0440}
\end{figure}

\begin{figure}
    \includegraphics[width=\columnwidth]{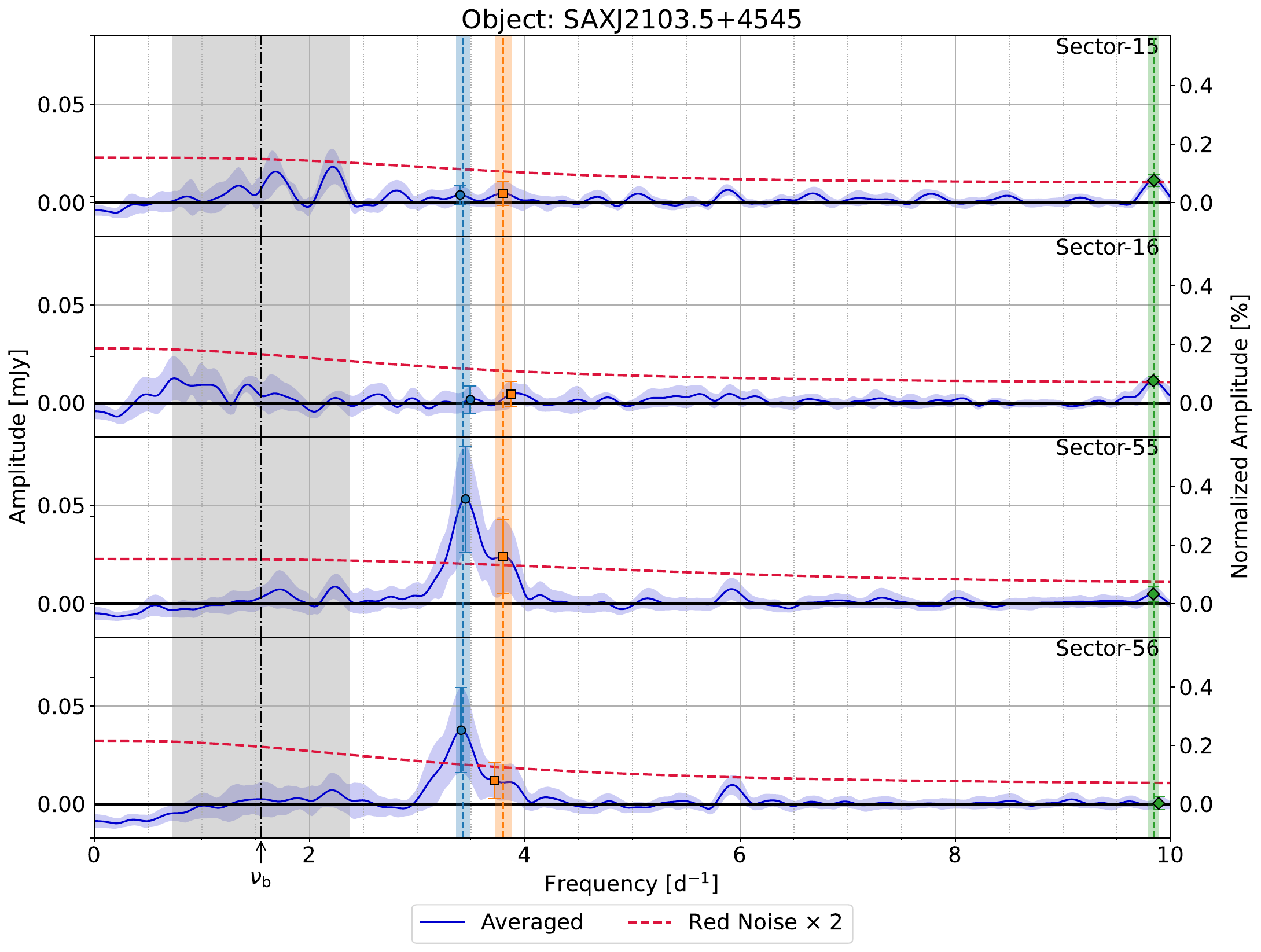}
    \caption{
        A similar figure to \cref{fig:spectra_4U0115} for SAX J2103.5+4545.
    }
    \label{fig:spectra_SAXJ2103}
\end{figure}

\begin{figure}
    \includegraphics[width=\columnwidth]{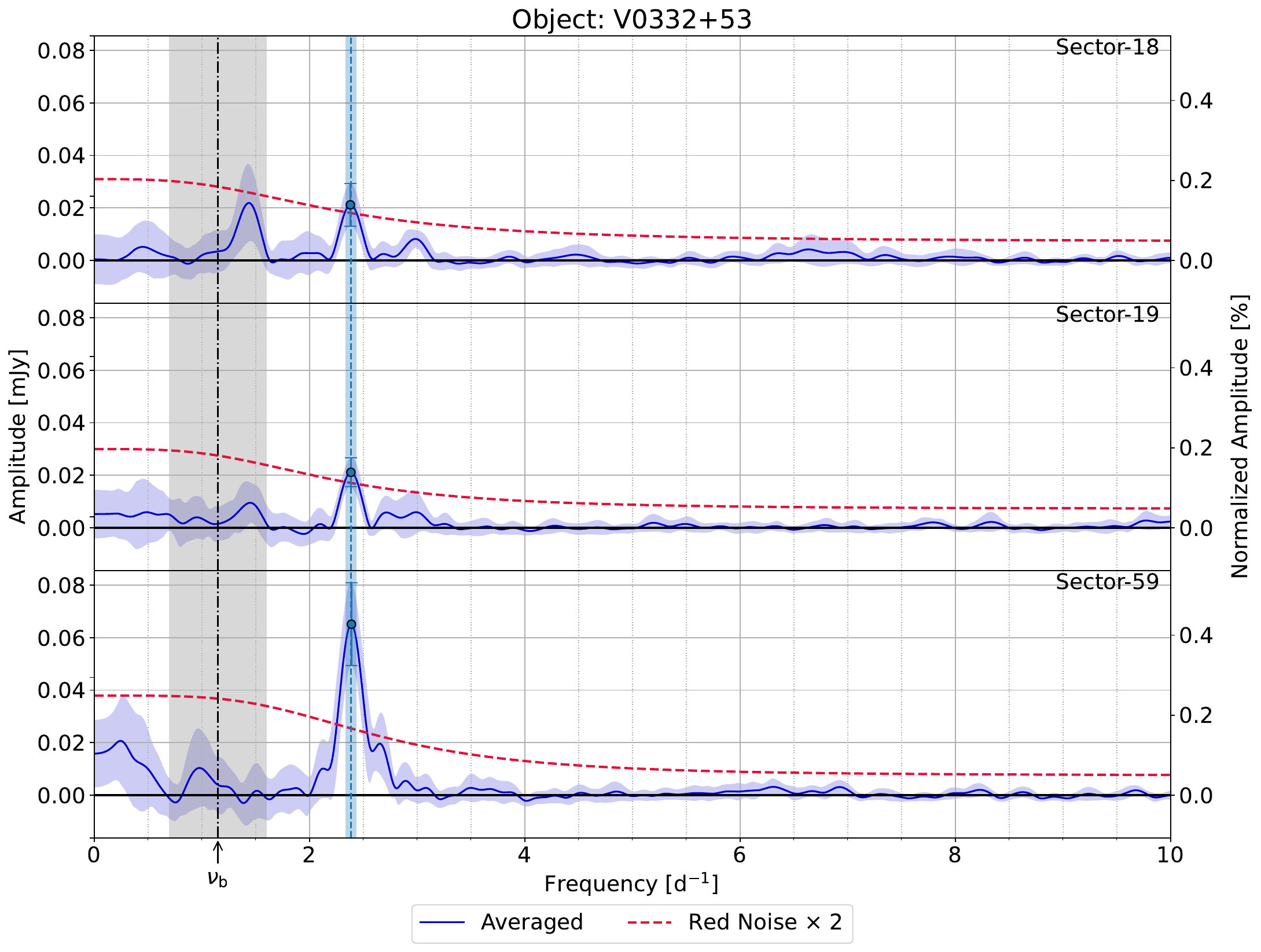}
    \caption{
        A similar figure to \cref{fig:spectra_4U0115} for V0332+53.
    }
    \label{fig:spectra_V0332}
\end{figure}

We summarised detected peak frequencies in \cref{tab:peaks}, and showed red noise subtracted amplitude spectra of five sources in \cref{fig:spectra_4U0115,fig:spectra_GROJ2058,fig:spectra_RXJ0440,fig:spectra_SAXJ2103,fig:spectra_V0332}.
When the ratio of two peak frequencies matched an integer within the $1\sigma$ error range, we determined that the pair of peaks has possible harmonic relationship, except in cases where the error of ratio exceeded 0.5.
The possible harmonic relationships were confirmed for only two of the five, 4U0115+634 and RX J0440.9+4431.
These were the only two having peaks of $\nu\lesssim\nub$.
In all five sources, the maximum peak was detected at $\nu=2\text{--}4~\dinv$ with a normalized amplitude of $\sim$ 1 per cent.
There was at least one peak per source where the normalized amplitude varied clearly between sectors.
And the maximum peak was the amplitude varying peak for all sources.

\begin{table}
    \caption{
        Detected peak frequencies of five sources.
        Peaks with a check mark in the `Harm.' column are those which have a possible integer multiple relationship to the frequency of the other peaks.
    }
    \begin{tabular}{ccc}
        \hline
        Name & $\nu~[\dinv]$ & Harm.\\
        \hline
        \multirow{5}{*}{4U0115+634} & $1.15\pm0.10$ & \checkmark\\
         & $1.63\pm0.11$ & \checkmark\\
         & $2.95\pm0.20$ & \\
         & $3.33\pm0.10$ & \checkmark\\
         & $6.65\pm0.10$ & \checkmark\\
        \hline
        \multirow{2}{*}{GRO J2058+42} & $2.37\pm0.10$ & \\
         & $6.38\pm0.10$ & \\
        \hline
        \multirow{5}{*}{RX J0440.9+4431} & $1.11\pm0.14$ & \checkmark\\
         & $1.28\pm0.11$ & \checkmark\\
         & $2.18\pm0.11$ & \checkmark\\
         & $2.59\pm0.10$ & \checkmark\\
         & $6.74\pm0.10$ & \checkmark\\
        \hline
        \multirow{3}{*}{SAX J2103.5+4545} & $3.43\pm0.14$ & \\
         & $3.80\pm0.16$ & \\
         & $9.84\pm0.10$ & \\
        \hline
        \multirow{1}{*}{V0332+53} & $2.39\pm0.10$ & \\
        \hline
    \end{tabular}
    \label{tab:peaks}
\end{table}

\subsection{Peak amplitude and the IR excess correlations}
\begin{figure}
    \includegraphics[width=\columnwidth]{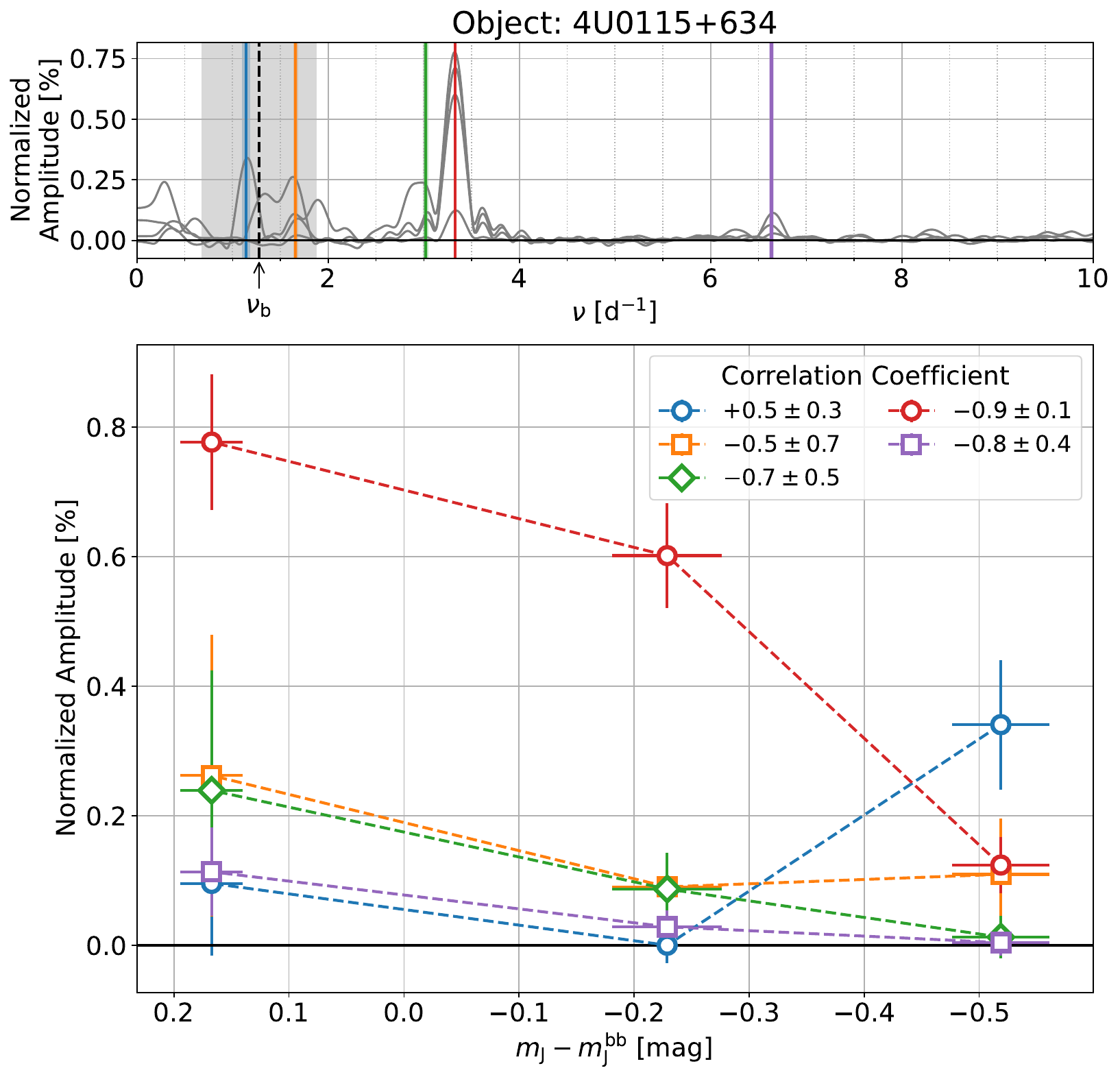}
    \caption{
        IR excess and peak amplitude relationship in 4U0115+634.
        (top) Red noise subtracted normalised amplitude spectrum.
        Basically the same as \cref{fig:spectra_4U0115}.
        (bottom) The further to the right, the larger the IR excess.
        Points with the same marker style and connected by dashed lines correspond to the peak of the same frequency and are plotted with values in multiple sectors.
        The color of the marker is the same as that of the vertical line in the top panel.
        In the upper right box are the correlation coefficients for each peak.
        Note that dashed lines simply connect adjacent points and ignore the time series.
    }
    \label{fig:corr_4U0115}
\end{figure}

\begin{figure}
    \includegraphics[width=\columnwidth]{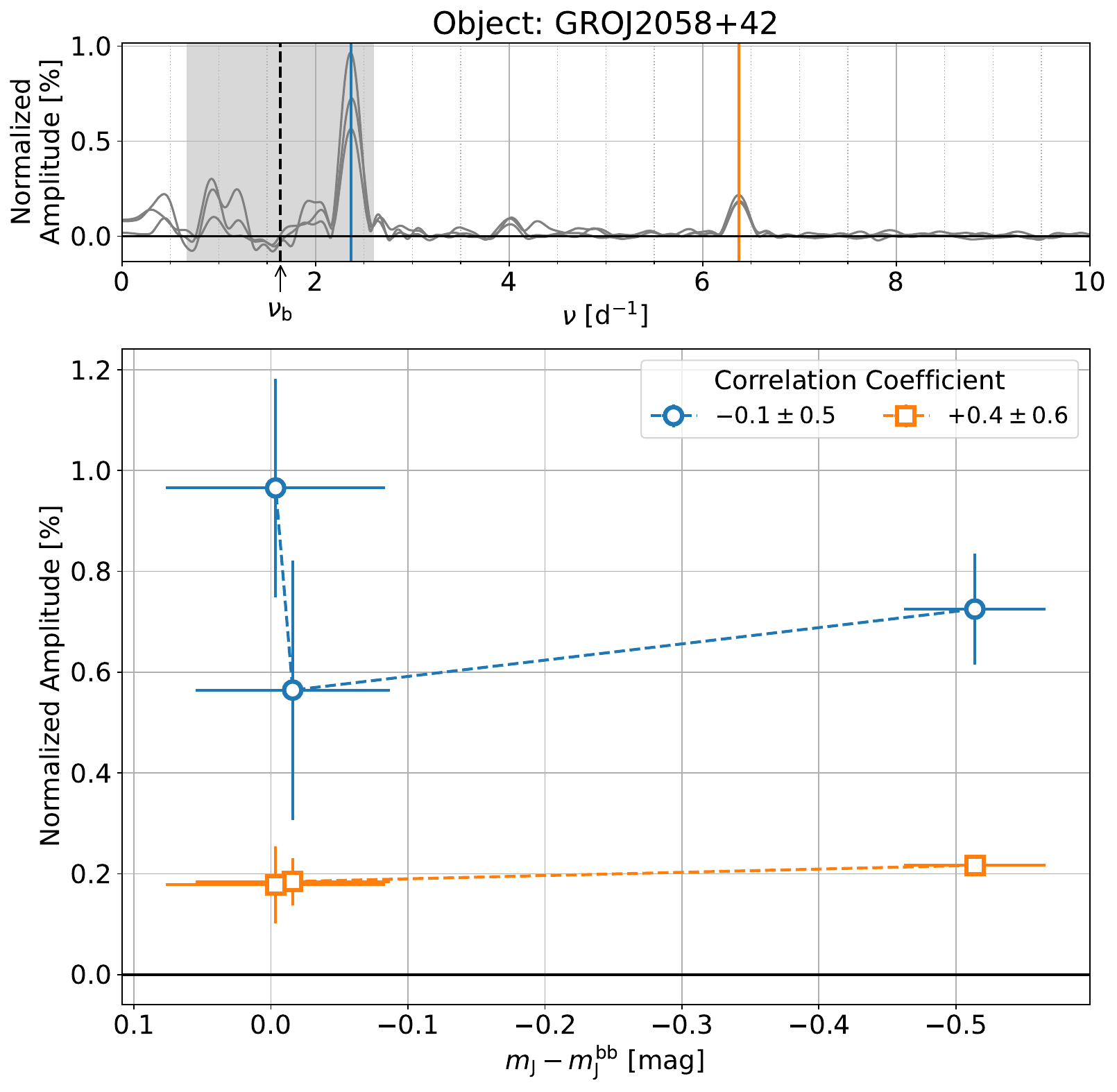}
    \caption{
        A similar figure to \cref{fig:corr_4U0115} for GRO J2058+42.
    }
    \label{fig:corr_GROJ2058}
\end{figure}

\begin{figure}
    \includegraphics[width=\columnwidth]{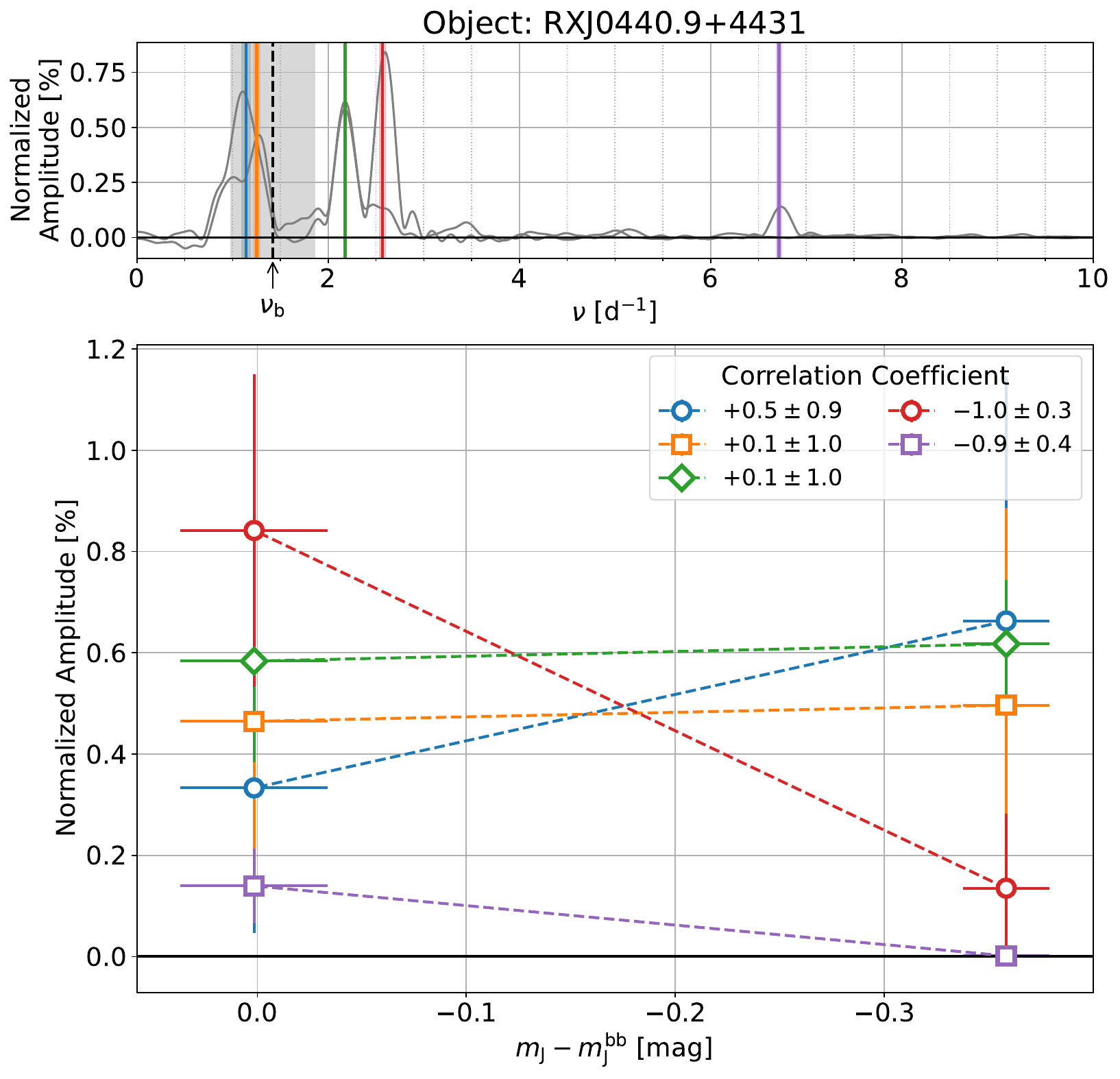}
    \caption{
        A similar figure to \cref{fig:corr_4U0115} for RXJ 0440.9+4431.
    }
    \label{fig:corr_RXJ0440}
\end{figure}

\begin{figure}
    \includegraphics[width=\columnwidth]{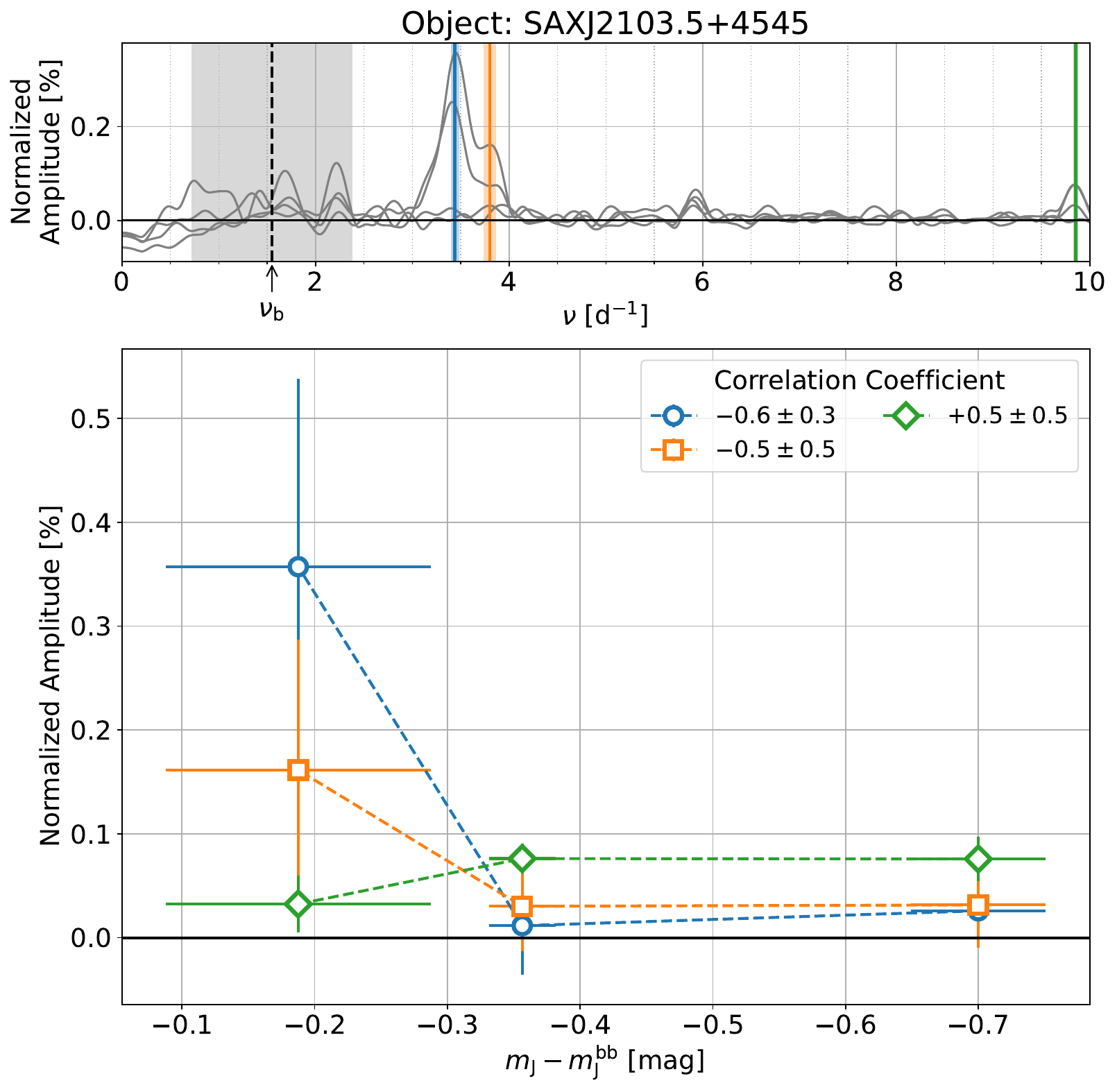}
    \caption{
        A similar figure to \cref{fig:corr_4U0115} for SAX J2103.5+4545.
    }
    \label{fig:corr_SAXJ2103}
\end{figure}

\begin{figure}
    \includegraphics[width=\columnwidth]{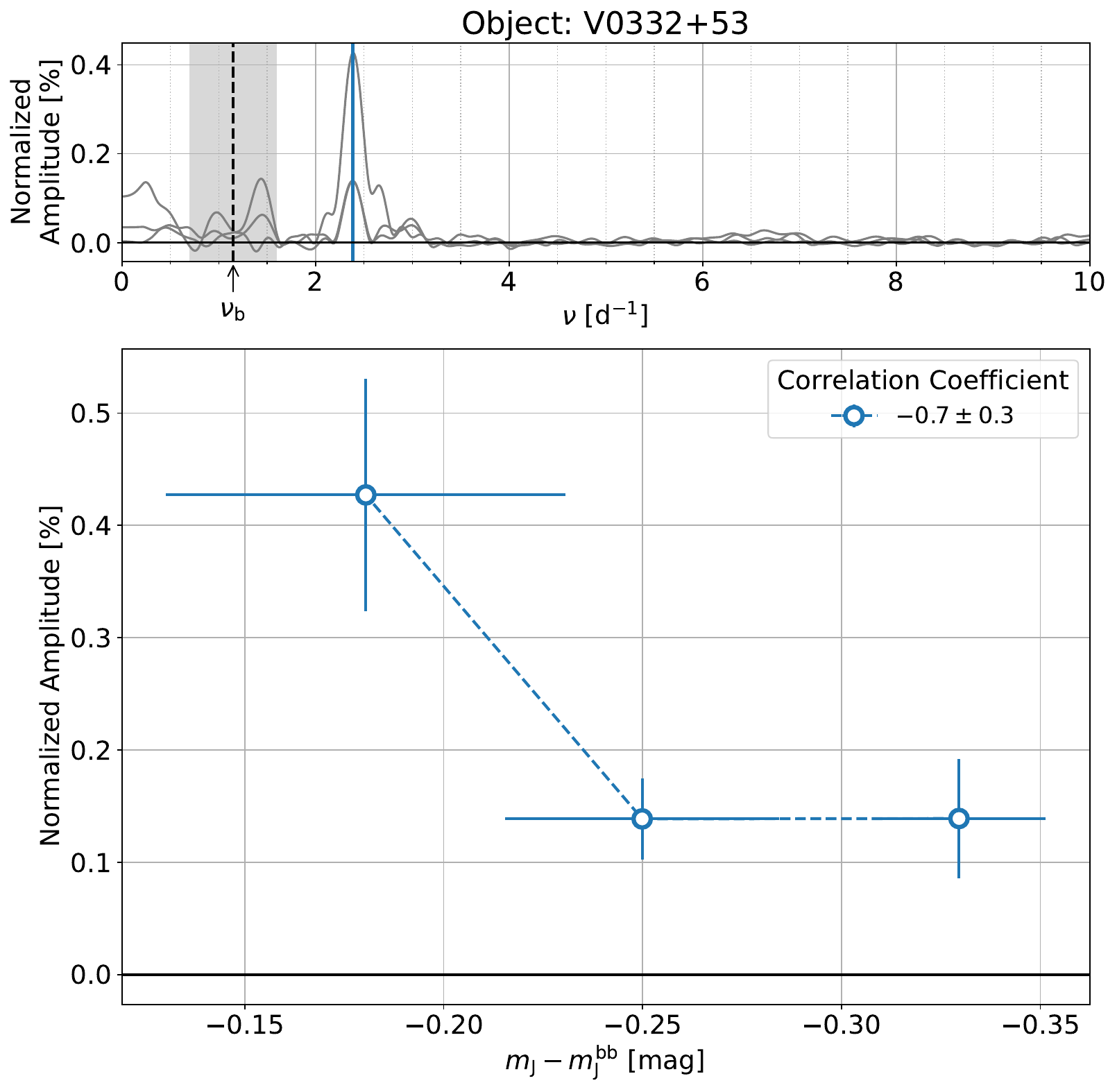}
    \caption{
        A similar figure to \cref{fig:corr_4U0115} for V0332+53.
    }
    \label{fig:corr_V0332+53}
\end{figure}

\cref{fig:corr_4U0115,fig:corr_GROJ2058,fig:corr_RXJ0440,fig:corr_SAXJ2103,fig:corr_V0332+53} show the relationship between peak amplitudes and the IR excess for five sources.
For all sources except GRO J2058+42, at least one peak exhibited significant anti-correlation between its amplitude and the IR excess.
Other peaks did not show significant correlations.
We refer these peaks as `anti-correlated peaks' and `non-correlated peaks', respectively.
Except for GRO J2058+42, the maximum peak belonged to the anti-correlated peaks.
The ratio of the maximum to minimum amplitudes of the anti-correlated peaks was typically 5--6.
For RX J0440.9+4431, both the anti-correlated and non-correlated peaks coexist (e.g. peaks at $\nu=2.2, 2.6$).

\section{Discussion}\label{sec:discussion}
\subsection{Origin of periodic flux oscillations}\label{sec:peak_clasiffication}
The origin of the periodic flux oscillation, which appears as a peak in the amplitude spectrum, is the rotation or pulsation of the Be donor star.
According to \citet{balona_2019} and \citet{balona_2020}, rotation-derived flux oscillations in massive stars are originated from gas clouds co-rotating with the star, and these appear in the power spectrum as clusters of peaks or broad humps, and integer multiplied harmonics.
Furthermore, the frequency of the rotation-derived fundamental mode should be smaller than $\nub$, because the rotation of the Be star does not reach a critical speed in most cases.
In light of this, peaks of $\nu<2~\dinv$ in 4U0115+634 and RX J0440.9+4431 may be rotation-derived fundamental modes, and their integer multiples may be harmonics.
On the other hand, the other peaks do not have rotation-derived features, and are considered to be pulsation-derived.
However, the possibility that they are harmonics of the rotation-derived peaks cannot be completely ruled out.
Furthermore, it should be noted that the ratio of the frequencies of the rotation-derived and pulsation-derived peaks could be close to an integer by chance, or that either the fundamental or harmonics could not be detected due to lack of signal-to-noise ratio.

\subsection{Interpretation of observed anti-correlations}\label{sec:anticorrelation}
The anti-correlation between flux oscillation amplitude and the IR excess is a notable result.
According to the conventional idea of the pulsation-driven mass ejection, the pulsations should be active when the disk is growing, and thus the flux oscillation amplitude should be positively correlated with the IR excess, but the result was the opposite.
In other words, if the anti-correlated peak amplitude variation we found is derived by the variation in the pulsation amplitude, it may disprove pulsation-driven mass ejection.
However, if the pulsation amplitude did not change but only the flux oscillation amplitude changed, or if the anti-correlated peak is not pulsation-derived but rotation-derived, it is possible to explain the result without denying the pulsation-driven mass ejection.
We examined following four scenarios to interpret the result:
\begin{description}
    \item[1st:] Co-rotating gas cloud scenario,
    \item[2nd:] Fully covered photosphere scenario,
    \item[3rd:] Partially covered photosphere scenario,
    \item[4th:] Internal state transition scenario.
\end{description}

We have to mention the possibility that peak amplitude variations were caused by low-frequency beats ($\nu\sim1/100~\dinv$), rather than the astrophysical mechanism.
However, in this case, the anti-correlation can only be explained as being due to observations made at phases that happen to be so, which is inconsistent with the fact that anti-correlations were confirmed in four of the five sources.
It is also possible that the anti-correlations are outliers.
Our data set has too small a sample size to examine statistical significance.

\subsubsection{1st: Co-rotating gas cloud scenario}\label{sec:scenario1}
\begin{figure}
    \includegraphics[width=\columnwidth]{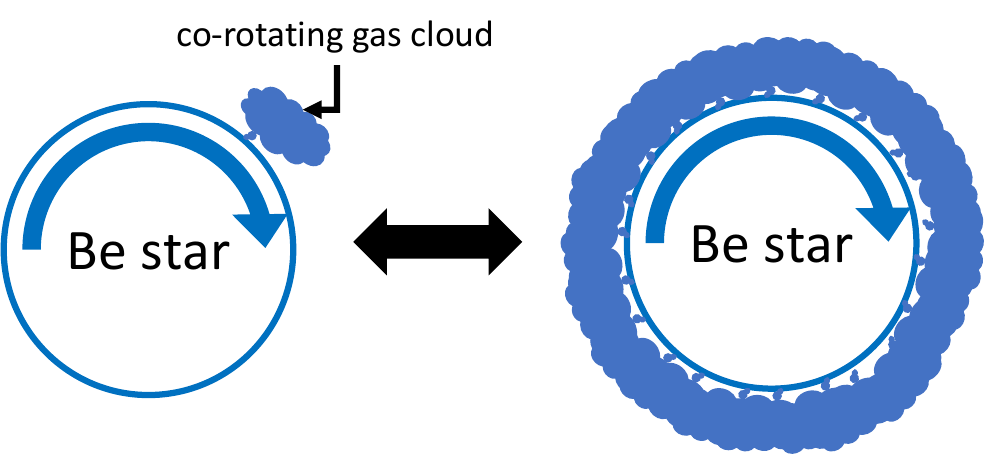}
    \caption{
        Schematics of the first scenario.
        These illustrations view the Be star along its rotation axis.
        On the left is the case of a low mass ejection, which has a larger dynamic range of co-rotating gas cloud density along the rotational direction.
        On the other hand, the right side is when mass ejection is active, and the density distribution in the rotational direction is homogenised by the increased amount of co-rotating gas, resulting in a smaller dynamic range of density.
    }
    \label{fig:scenario1}
\end{figure}

In the first scenario, we consider that changes in the density distribution of co-rotating gas clouds are the cause of amplitude variations.
In other words, the origin of flux oscillations are not pulsations but the rotation.
Since rotation-derived flux oscillations are caused by co-rotating gas clouds, the amplitude of flux oscillation should be determined by the dynamic range of gas cloud density around the rotation axis.
Therefore, the peak amplitude is expected to decrease in conjunction with the mass ejection of the Be star and the disk growth due to that the amount of co-rotating gas increased and homogenised around the rotation axis (\cref{fig:scenario1}).
However, unlike pulsation, which can have multiple modes, rotation is a single mode oscillation, so it is unlikely that rotation-derived anti-correlated and non-correlated peaks coexist.
Furthermore, this scenario provides no explanation for amplitude variations of pulsation-derived flux oscillations.
Therefore, in this scenario, all anti-correlated and non-correlated peaks must be rotation-derived and pulsation-derived, respectively.
Nevertheless, only 4U0115+634 and RX J0440.9+4431 have peaks with features suggesting a rotation origin, and to apply this scenario to other sources, it would have to be that the harmonics were detected but the fundamental mode was not.

\subsubsection{2nd: Fully covered photosphere scenario}\label{sec:scenario2}
\begin{figure}
    \includegraphics[width=\columnwidth]{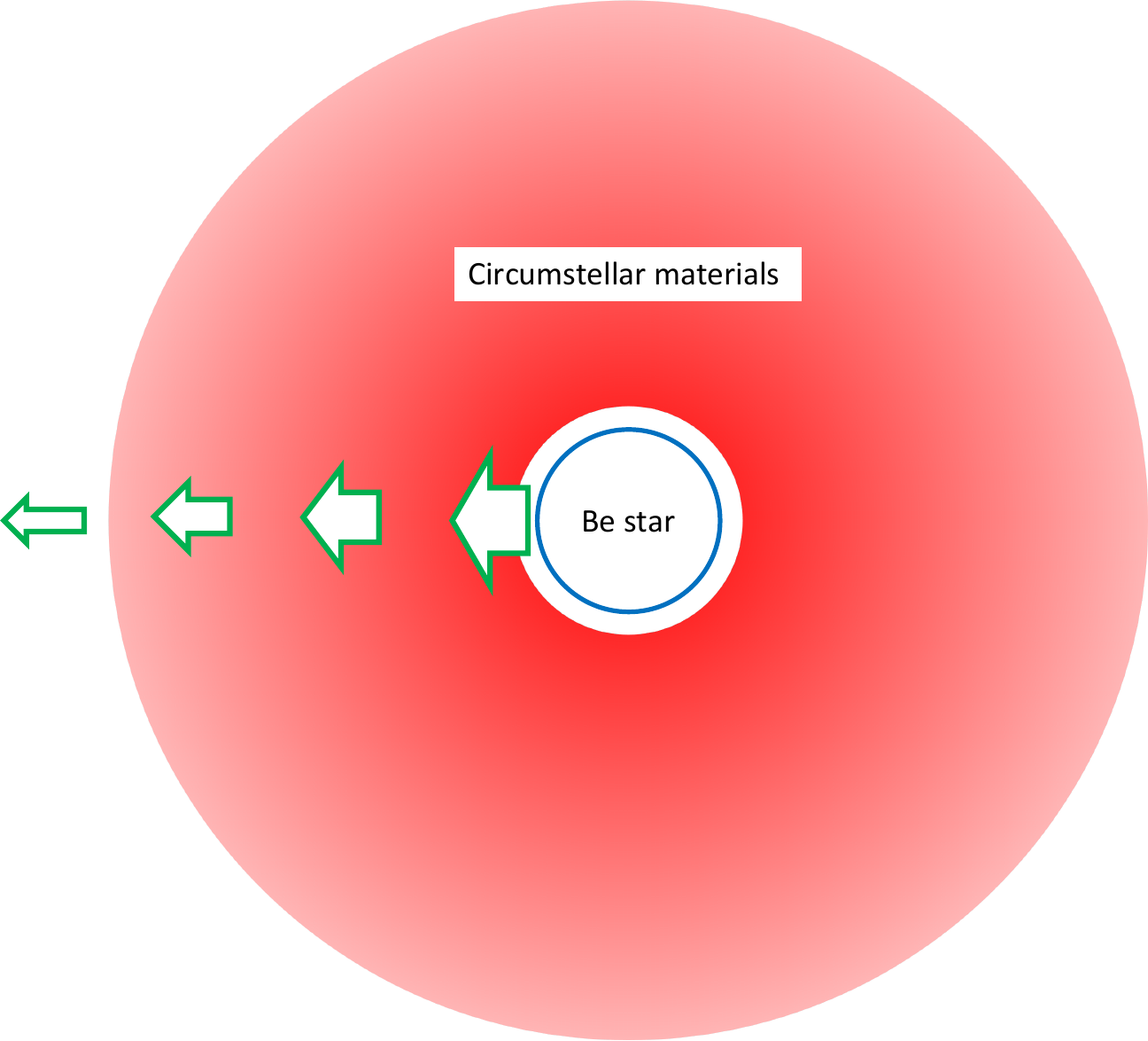}
    \caption{
        Schematics of the second scenario.
        The green arrows are photon streams from the Be star, and the size indicates the number of photons pass through without scattering.
        In this scenario, the photons coming from the Be star are uniformly scattered by surrounding circumstellar materials at the same rate.
    }
    \label{fig:scenario2}
\end{figure}

The second scenario is based on the idea that the amplitude of the pulsation itself are constant, but the flux oscillation amplitude varies linked to the change in the amount of circumstellar materials (e.g. disk, stellar wind) and their reprocessing.
We considers a situation in which most or all of the photosphere is covered by the circumstellar materials, as seen from our perspective.
In this scenario, the photons coming from the Be star are scattered by the circumstellar materials, with at least 20 per cent of the photons passing through without being scattered even once (\cref{fig:scenario2}).
If the main scattering process is Thomson scattering, the reprocessing itself does not affect on the SED.
However, because photons emitted from various regions on the photosphere are mixed, the scattered flux is considered to have lost periodicity.
Since the mass ejected from the Be star should be supplied not only to the disk but also to 
the circumstellar materials, we expect to observe the peak amplitude decrease in conjunction with the disk growth.
This logic can also be applied to explain rotation-derived oscillations.
However, all peak amplitudes should decrease uniformly because almost all photons are reprocessed at the same rate in this mechanism.
Thus, this scenario cannot explain non-correlated peaks.

\begin{figure}
    \includegraphics[width=\columnwidth]{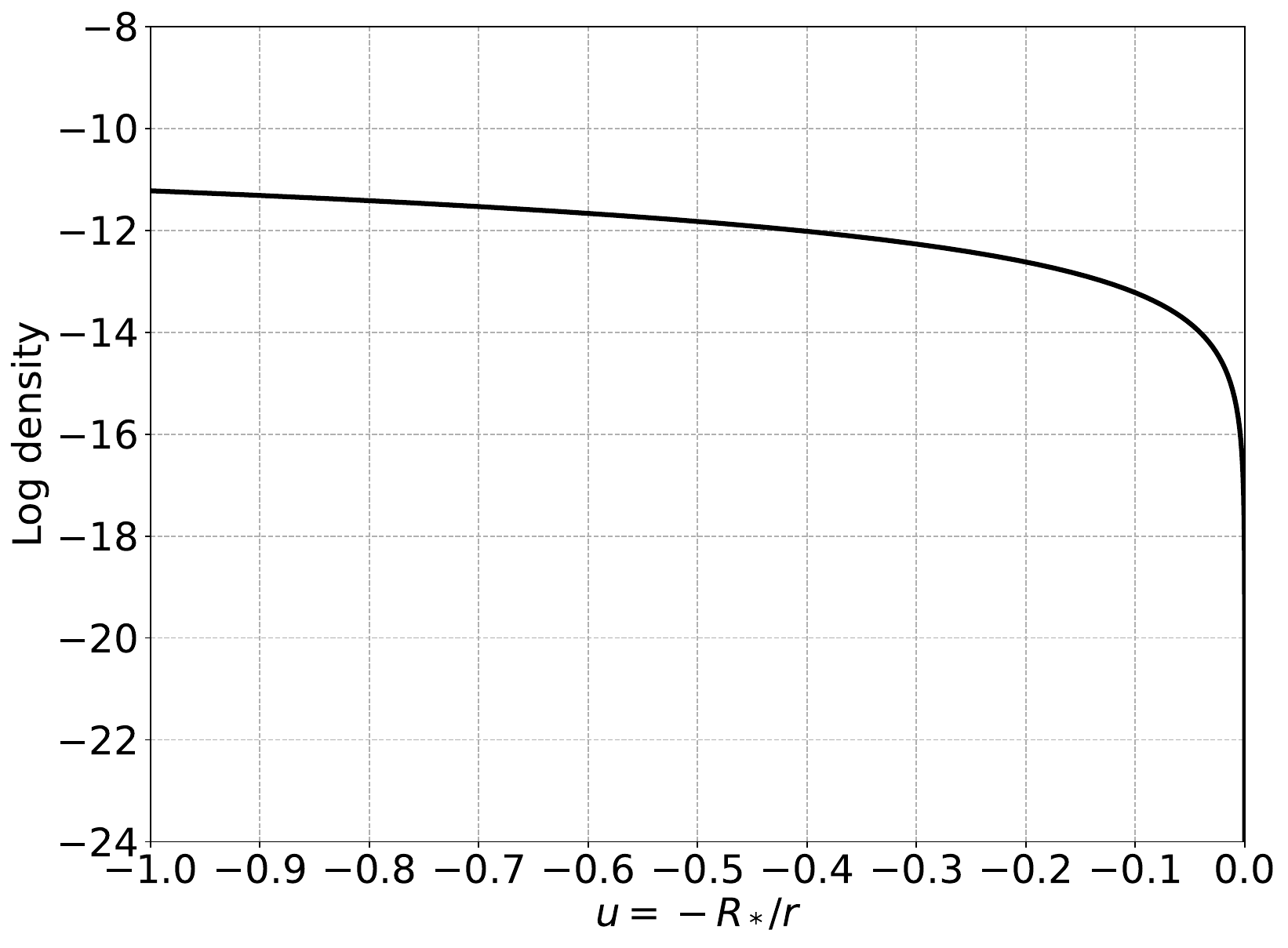}
    \caption{
        Roughly estimated density profile.
        The unit of density is $\gpqcm$ and the base of logarithm is 10.
        The flatness in $u\sim-1$ is due to the neglect of particle acceleration near the stellar surface.
        The range of the vertical axis is the same as in Figure 15 of \citet{cure_2004}.
    }
    \label{fig:densityprofile}
\end{figure}

Because the situation in this scenario is simple, it is easy to estimate the density of 
the circumstellar materials.
The following relationship is expected to be satisfied when photons pass through a material:
\begin{equation}
    \tau = \napier^{-N\sigma},\label{eq:transmittance}
\end{equation}
where $\tau$ is a transmittance, $N$ is a electron column density, and $\sigma$ is a scattering cross section.
Substituting $\tau=0.2$ and the Thomson scattering cross section $\sigma_T=6.7\times10^{-25}~\mathrm{cm^2}$ for $\sigma$, the column density is calculated as $N=2.4\times10^{24}~\mathrm{cm^{-2}}$.
Assuming a spherical symmetric stellar wind as scattering materials, an electron number density $n$ is expected to follow an inverse square law with respect to a distance from the star $r$ due to the conservation of particle number, when ignoring particle acceleration:
\begin{equation}
    n = n_0\brkts{\frac{R*}{r}}^2,\label{eq:number_distribution}
\end{equation}
where $n_0$ is a number density at the surface and $\Rstar$ is a radius of the star.
$N$ is then equal to integral of $n$ from the stellar surface to infinity:
\begin{align}
    N =& \int_{\Rstar}^{\infty}n~\diff{r}\\
     =& n_0\Rstar.\label{eq:column_density}
\end{align}
If $\Rstar=10\Rsun$, then $n_0=3.4\times10^{12}~\mathrm{cm^{-3}}$.
Assuming that the material is hydrogen, the numbers of electrons and nucleons are equal, thus a density at the surface is $\rho=6\times10^{-12}~\gpqcm$.
This is comparable to the observed density of the inner disk ($=10^{-13}\text{--}10^{-10}~\gpqcm$, cf. \citealp{gies_2007,pott_2010,schaefer_2010,stefl_2012,meilland_2012}).
\cref{fig:densityprofile} shows the estimated density profile.
The estimated density is $\sim$ 100 times larger than the numerical calculated wind density profile of the B1Ve star by \citet{cure_2004}.
If the density distribution is not spherically symmetric, i.e., if the circumstellar material in our line of sight is locally dense, the less average density can be accounted for.
In this case, there should be a clump at least 100 times denser than normal spherical symmetric winds.
For example, the winds of Wolf-Rayet (WR) stars are clumpy as indicated by the stochastic variation in the line profiles \citep{moffat_1988}.
The flux-limiting radiation hydrodynamic simulation of \citet{moens_2022} predicted a dynamic range of $\sim100$ in density of clumpy winds of WR stars.
In addition, analysis of the UV and optical spectra of eight galactic O-type supergiants by \citet{hawcroft_2021} showed that half of the wind velocity field was covered by dense clumps and that their density was 25 times the average density.

Incidentally, it is not obvious whether the effect of the temperature increase of the photosphere due to the return of some of the scattered photons to the Be star can be ignored in the situation where 80 per cent of the photons are scattered.
So, we calculated the fraction of returning photons and obtained the value of about 10--20 per cent in the case considered here (\cref{apd:photon_fraction}).
Thus, at least the effect on the estimation of Be star properties in \cref{sec:be_property} can be considered negligible.

\subsubsection{3rd: Partially covered photosphere scenario}\label{sec:scenario3}
\begin{figure}
    \includegraphics[width=\columnwidth]{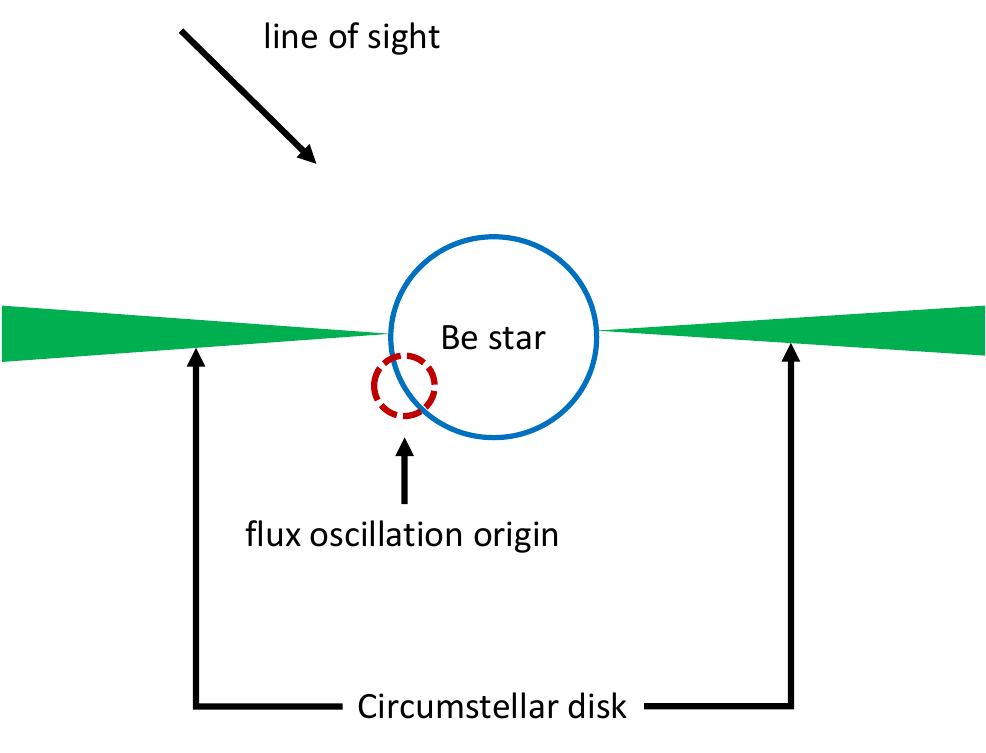}
    \caption{
        Schematics of the third scenario.
        In this scenario, the flux oscillation amplitude is reduced because the region near the Be star or the photosphere contributing to the flux oscillation is covered by something around the Be star (in this figure, the disk) when viewed from our perspective.
        If the flux oscillations are caused by a larger region, this scenario cannot account for the amplitude variation of $\sim80~\mathrm{per cent}$.
    }
    \label{fig:scenario3}
\end{figure}

The third scenario shares the basic idea with the second one (\cref{sec:scenario2}) that 
the circumstellar materials obscure the Be star, but differs in that they obscure only a portion of the star, not the entire star (\cref{fig:scenario3}).
In this case, since some areas on the photosphere are not covered, the presence of non-correlated peaks can be tolerated.
The question here is whether it is possible to reduce the peak amplitude by 80 per cent just by hiding part of the photosphere.
For example, assuming the obscuration by the disk as shown in \cref{fig:scenario3}, the obscured area is at most 50 per cent of the entire photosphere, and it seems difficult to achieve the amplitude reduction of 80 per cent.
It can be satisfied in the case where the flux oscillations are generated in a limited region around the Be star.
For the rotation-derived flux oscillations, this logic is applicable because they originate from co-rotating gas clouds.
However, the probability that the co-rotating gas cloud is hidden by the disk is at most 50 per cent, so it is slightly difficult to explain why significant anti-correlations were found in four of the five sources.
On the other hand, it is non-trivial whether the requirement is satisfied for the pulsation-derived flux oscillations.
Non-radial pulsations cause the local expansion and contraction, and the geometry of the oscillation is characterised by the spherical harmonics.
The local brightness at the expanding region is expected to be smaller due to gravitational darkening.
In low-$l$ modes, the number of nodal surfaces appearing on the photosphere is less, so each area of expansion or contraction is relatively large, and it is thought that the expanding and contracting regions do not appear equal when viewed from one direction.
In this case, periodic flux oscillations can be generated by the brightness variations mentioned above, and the area contributing to the flux oscillations is relatively large.
On the other hand, the same logic cannot simply be applied to cases where the expanding and contracting regions appear equal, such as in high-$l$ modes.
In this case, flux oscillations can be generated by a local brightness difference (e.g. star spots) between the expanding and contracting regions.
Therefore, some of the pulsation-derived flux oscillations may be caused by a limited region on the photosphere.
However, the amplitude of the flux oscillations produced by this mechanism is considered to be smaller than that produced by the expansion and contraction of a larger region in low-$l$ modes.
That makes difficult to interpret the anti-correlations of the maximum peaks in this scenario.
In addition, local brightness differences on the photosphere are expected to be visible and hidden as the Be star rotates, which would cause rotation-linked features in light curves.
Therefore, it is also difficult to apply this scenario to SAX J2103.5+4545 and V0332+53, which have no significant rotation-derived features in their amplitude spectra.

\subsubsection{4th: Internal state transition scenario}\label{sec:scenario4}
\begin{figure}
    \includegraphics[width=\columnwidth]{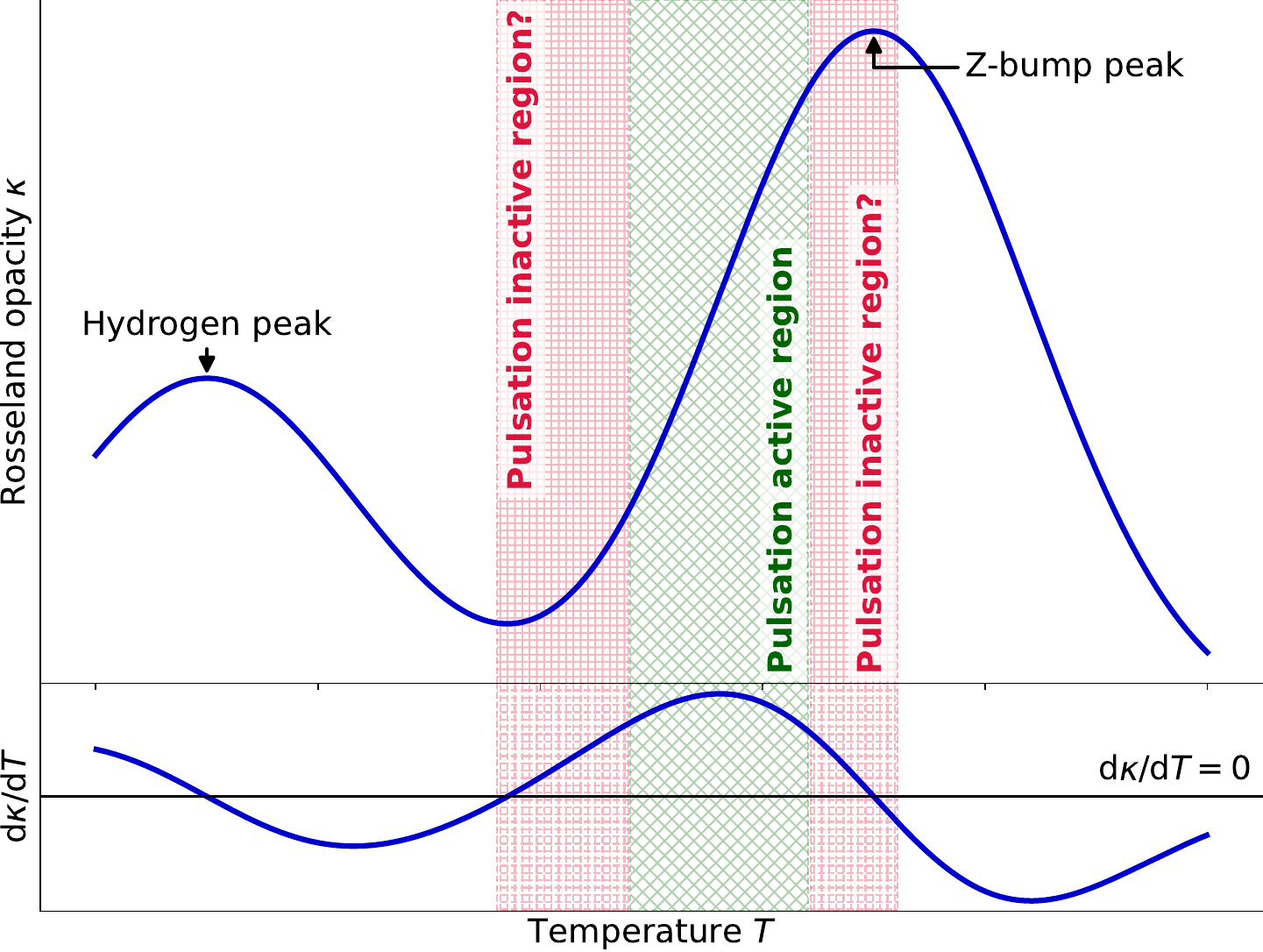}
    \caption{
        Schematic $T\text{-}\kappa$ curve and temperature dependence of $\mathrm{d}\kappa/\mathrm{d}T$.
        Pulsations should be most active where $\mathrm{d}\kappa/\mathrm{d}T$ is largest.
        And it is qualitatively expected that the pulsation will be inactive at higher or lower temperature regions.
    }
    \label{fig:scenario4}
\end{figure}

In the fourth scenario, we considered that the pulsation amplitude is indeed inversely correlated with the disk growth.
The pulsations in early-type massive stars are driven by the $\kappa$-mechanism \citep{dziembowski_1993a,dziembowski_1993b,gautschy_1993,pamyatnykh_1999,miglio_2007}.
In the temperature range where the Rosseland opacity $\kappa$ is positively correlated with the temperature $T$, $\kappa$ variations provide positive feedback to $T$ variations, causing growth of thermodynamic fluctuations and thus pulsations.
Since the amount of this feedback should depend on the slope of the $T\text{-}\kappa$ curve (i.e. $\mathrm{d}\kappa/\mathrm{d}T$), the intensity of $\kappa$-mechanism, and thus the pulsation amplitude, is expected to have dependence on the inner temperature distribution.
Therefore, we first assume that some variation in the temperature inside the Be star has caused the pulsation to become inactive.
The $\kappa$-mechanism is expected to be most active at the steepest point on the low-temperature side slope of the iron-group element derived bump (so-called Z-bump) in the $T\text{-}\kappa$ curve.
And therefore, $\mathrm{d}\kappa/\mathrm{d}T$ becomes smaller in the higher temperature as it is closer to the Z-bump peak.
Conversely, moving to the low temperature side also reduces $\mathrm{d}\kappa/\mathrm{d}T$ as it approaches the bottom of valley between Z-bump and the hydrogen-derived bump.
In other words, a decrease in pulsation amplitude is qualitatively expected for both temperature increases and decreases (\cref{fig:scenario4}).
If the star is hotter in the pulsation-inactive state than the pulsation-active state, relatively strong radiation pressure may promote mass ejection.
On the other hand, considering the low temperature pulsation-inactive state, the lower temperature may be caused by the star expansion and the outer layer is more loosely bound than the active state, which is also expected to be relatively favourable for mass ejection.
In any case, the pulsation-inactive state is qualitatively expected to be more active in mass ejection than the pulsation-active state.
Note that these are only qualitative discussions, and a more quantitative discussion requires a detailed understanding of the structure and temperature distribution inside Be stars.

Since this mechanism works for all pulsations, it cannot account for pulsation-derived non-correlated peaks.
In addition, this scenario provides no explanation for the rotation-derived flux oscillations.
Therefore, the origin of the anti-correlated and non-correlated peaks should be pulsation and the rotation, respectively.
This is not so unreasonable since most of the anti-correlated peaks did not have the rotation-derived features, as discussed in \cref{sec:peak_clasiffication}.
\section{Conclusions}\label{sec:conclusion}
The BeXBs exhibit characteristic and poorly understood phenomena such as two types of outbursts.
To clarify them, we need to understand the nature of the circumstellar disk of the Be donor star.
In particular, the mechanism driving the mass ejection of Be stars, which grows their circumstellar disks, is one of the critical unresolved issues.

We analysed long-term variations in the flux periodicity of five galactic BeXBs using $\tess$ light curves, and compared them with the long-term multi-wavelength light curves, to constrain the behaviour of their circumstellar disk.
We obtained X-ray, optical, and NIR light curves for $\swift$, MAXI, ZTF, ATLAS, and Gattini-IR.
Then we confirmed that there was OIR variability on time scale of hundreds of days derived from the circumstellar disk and they exhibited giant outbursts.
We extrapolated the $J$-band magnitudes from the optical catalogued magnitudes using a simple blackbody model and subtracted them from the Gattini-IR $J$-band light curve to evaluate the IR excess.
In addition, we made amplitude spectra from $\tess$ light curves, and extract peaks of periodic flux oscillations derived by the rotation or pulsations.

As a result, we found the anti-correlations between peak amplitudes and the IR excess.
We examined the four scenarios to explain these results.
\cref{tab:scenarios} shows whether each scenario can explain the observed results.
\begin{table}
    \caption{
        Summary of four scenarios.
        It shows whether each scenario can explain the observed results.
        `$\bigcirc$', `$\times$', and `$\bigtriangleup$' indicate explainable, inexplicable, and conditionally explainable, respectively.
    }
    \begin{tabular}{cccccc}
        \hline
        \multicolumn{2}{c}{Observed features} & \rotatebox{90}{Co-rotating}\rotatebox{90}{gas cloud} & \rotatebox{90}{Fully covered}\rotatebox{90}{photosphere} & \rotatebox{90}{Partially covered}\rotatebox{90}{photosphere} & \rotatebox{90}{Internal}\rotatebox{90}{state transition}\\
         \hline
        \multirow{2}{*}{Anti-correlated peaks} & Rotation & $\bigcirc$ & $\bigtriangleup$ & $\bigtriangleup$ & $\times$\\
        & Pulsation & $\times$ & $\bigtriangleup$ & $\bigtriangleup$ & $\bigcirc$\\
        \multirow{2}{*}{Non-correlated peaks} & Rotation & $\times$ & $\times$ & $\bigcirc$ & $\bigcirc$\\
        & Pulsation & $\bigcirc$ & $\times$ & $\bigcirc$ & $\times$\\
        \multicolumn{2}{c}{Anti-correlation of maximum peaks} & $\bigcirc$ & $\bigtriangleup$ & $\times$ & $\bigcirc$\\
        \multicolumn{2}{c}{Absence of rotation-derived peaks} & $\times$ & $\bigcirc$ & $\times$ & $\bigcirc$\\
        \hline
    \end{tabular}
    \label{tab:scenarios}
\end{table}
The first scenario, which assumes that all pulsation-derived oscillations appears as non-correlated peaks, is unreasonable, and the second and third scenarios, which attempt to explain anti-correlations by reprocessing of the circumstellar materials, have severe conditions to explain the results.
Therefore, we determined that it is most reasonable to interpret the observed anti-correlations as indicating an true anti-correlation between pulsation amplitudes and disk growth.
In other words, the conventional idea of pulsation-driven mass ejection of Be stars may be incorrect.
From this perspective, the most plausible scenario is the fourth one that assumed Be star internal state transition.

There were only two or three sectors with both $\tess$ and Gattini-IR data available for each source, so it is difficult to examine the statistical significance of the confirmed anti-correlations.
In addition, there were no $\tess$ observations covering the near-infrared brightening phase when the mass ejection should be most active, for the target sources of this study.
Therefore, further accumulation of data from future $\tess$ observations is desirable.
It is also important to check whether other BeXBs exhibit the same anti-correlation.

\section*{Acknowledgements}
M. Niwano was supported by Grant-in-Aid for JSPS Fellows.
This research was supported by JSPS KAKENHI Grant Number 23KJ0913, and 23K25910.
This work includes data collected by the $\tess$ mission, and funding for the $\tess$ mission is provided by NASA's Science Mission Directorate.
This research has made use of the MAXI data provided by RIKEN, JAXA, and the MAXI team.
We acknowledge the Samuel Oschin 48-inch Telescope at the Palomar Observatory as part of the Zwicky Transient Facility project, supported by the National Science Foundation under Grant No. AST-1440341 and a collaboration including Caltech, IPAC, the Weizmann Institute for Science, the Oskar Klein Center at Stockholm University, the University of Maryland, the University of Washington, Deutsches Elektronen-Synchrotron, and Humboldt University, Los Alamos National Laboratories, the TANGO Consortium of Taiwan, the University of Wisconsin at Milwaukee, and Lawrence Berkeley National Laboratories. Operations are conducted by COO, IPAC, and UW.
This work has made use of data from the Asteroid Terrestrial-impact Last Alert System (ATLAS) project. The Asteroid Terrestrial-impact Last Alert System (ATLAS) project is primarily funded to search for near earth asteroids through NASA grants NN12AR55G, 80NSSC18K0284, and 80NSSC18K1575; byproducts of the NEO search include images and catalogs from the survey area. This work was partially funded by Kepler/K2 grant J1944/80NSSC19K0112 and HST GO-15889, and STFC grants ST/T000198/1 and ST/S006109/1. The ATLAS science products have been made possible through the contributions of the University of Hawaii Institute for Astronomy, the Queen’s University Belfast, the Space Telescope Science Institute, the South African Astronomical Observatory, and The Millennium Institute of Astrophysics (MAS), Chile.
This paper makes use of data from the AAVSO Photometric All Sky Survey, funded by the Robert Martin Ayers Sciences Fund and NSF (AST-1412587).
The Pan-STARRS1 Surveys (PS1) and the PS1 public science archive have been made possible through contributions by the Institute for Astronomy, the University of Hawaii, the Pan-STARRS Project Office, the Max-Planck Society and its participating institutes, the Max Planck Institute for Astronomy, Heidelberg, and the Max Planck Institute for Extraterrestrial Physics, Garching, The Johns Hopkins University, Durham University, the University of Edinburgh, the Queen's University Belfast, the Harvard-Smithsonian Center for Astrophysics, the Las Cumbres Observatory Global Telescope Network Incorporated, the National Central University of Taiwan, the Space Telescope Science Institute, the National Aeronautics and Space Administration under Grant No. NNX08AR22G issued through the Planetary Science Division of the NASA Science Mission Directorate, the National Science Foundation Grant No. AST-1238877, the University of Maryland, Eotvos Lorand University (ELTE), the Los Alamos National Laboratory, and the Gordon and Betty Moore Foundation.
This work has made use of data from the European Space Agency (ESA) mission Gaia (\url{https://www.cosmos.esa.int/gaia}), processed by the Gaia Data Processing and Analysis Consortium (DPAC, \url{https://www.cosmos.esa.int/web/gaia/dpac/consortium}). Funding for the DPAC has been provided by national institutions, in particular the institutions participating in the Gaia Multilateral Agreement.
This work utilized Python libraries for analysis and visualization: NumPy \citep{harris_2020}, SciPy \citep{virtanen_2020}, pandas \citep{mckinney_2010}, Matplotlib \citep{hunter_2007}.
This work made use of Astropy:\footnote{\url{http://www.astropy.org}} a community-developed core Python package and an ecosystem of tools and resources for astronomy \citep{astropy_2013, astropy_2018, astropy_2022}.

\section*{Data Availability}
$\swift$/BAT and ZTF light curves used in this work are available in web pages of $\swift$/BAT Hard X-ray Transient Monitor\footnote{\url{https://swift.gsfc.nasa.gov/results/transients/}} and NASA/IPAC Infrared Science Archive\footnote{{\url{https://irsa.ipac.caltech.edu/Missions/ztf.html}}}, respectively.
ZTF data we used are included in the ZTF public data release 15.
The MAXI and ATLAS light curves in this work were generated by publicly available web services: MAXI on-demand\footnote{\url{http://maxi.riken.jp/mxondem/}} and ATLAS Forced Photometry\footnote{\url{https://fallingstar-data.com/forcedphot/}}, respectively.
As for the $\tess$ light curves, the FFIs are available from MAST\footnote{\url{https://archive.stsci.edu/missions-and-data/tess}}.
The Gattini-IR light curve is not publicly available.



\bibliographystyle{mnras}
\bibliography{
    bibfiles/bib_1900,
    bibfiles/bib_2000,
    bibfiles/bib_2005,
    bibfiles/bib_2010,
    bibfiles/bib_2015,
    bibfiles/bib_2020
}




\appendix
\section{Typical properties of O--B type main-sequence stars}\label{apd:main_sequence}
\cref{tab:spt} summaries typical properties of O--B type main sequence stars \citep{pecaut_2013}.

\begin{table}
    \caption{
        Typical properties of O--B type main sequence stars \citep{pecaut_2013}.
        Extracted from \url{https://www.pas.rochester.edu/~emamajek/EEM_dwarf_UBVIJHK_colors_Teff.txt}.
    }
    \begin{tabular}{cccc}
        \hline
        SpT & $\Teff$ & $\Rstar$ & $\Mstar$\\
         & [kK] & [$\Rsun$] & [$\Msun$]\\
        \hline
        O7.5V & 36.1 & 8.95 & 26.0\\
        O8V & 35.1 & 8.47 & 23.6\\
        O8.5V & 34.3 & 8.06 & 21.9\\
        O9V & 33.3 & 7.72 & 20.2\\
        O9.5V & 31.9 & 7.50 & 18.7\\
        B0V & 31.4 & 7.16 & 17.7\\
        B0.5V & 29.0 & 6.48 & 14.8\\
        B1V & 26.0 & 5.71 & 11.8\\
        B1.5V & 24.5 & 5.02 & 9.9\\
        B2V & 20.6 & 4.06 & 7.3\\
        \hline
    \end{tabular}
    \label{tab:spt}
\end{table}

\section{A fraction of photons returning to the Be star due to scattering in spherically symmetric stellar wind}\label{apd:photon_fraction}
A variation in the number of photons $\Nphoton$ as they pass through matter over a microscopic distance $\diff{r}$ can be written as follows:
\begin{equation}
    \diff{\Nphoton} = -n\sigma\Nphoton\diff{r},\label{eq:differential_photon_number}
\end{equation}
where $n$ is a electron number density and $\sigma$ is a scattering cross section.
By solving \cref{eq:differential_photon_number} with substituting \cref{eq:number_distribution}, a number of pass-through photons $\Npass$ is obtained:
\begin{align}
    \Npass &= \Nphoton_0\exp{\brktl{-n_0\sigma\Rstar\brkts{1-\frac{\Rstar}{r}}}},\\
    &= \Nphoton_0\exp{\brktl{-N\sigma\brkts{1-\tilde{r}^{-1}}}}, & \brkts{N=n_0\Rstar,~ \tilde{r}=r/\Rstar}\label{eq:passing_photons}\\
    &= \Nphoton_0\exp{\brktl{-A\brkts{1-\tilde{r}^{-1}}}}, & \brkts{A\equiv N\sigma}
\end{align}
where $\Rstar$ is a stellar radius, $\Nphoton_0$ and $n_0$ are a number of photons and an electron number density at the stellar surface ($r=\Rstar$), $N$ is an electron column density, and $\tilde{r}$ is a normalised radius.
Note that \cref{eq:passing_photons} is consistent with \cref{eq:transmittance} in the limit of $\tilde{r}\rightarrow\infty$.
In terms of the model in \cref{sec:scenario2}, $\Npass$ corresponds to the number of photons passing a sphere of radius $\tilde{r}$ that have never experienced scattering.
Since scattering experienced photons also pass through the same plane, $\Nphoton$ is expected to be between $\Npass$ and $\Nphoton_0$.
The differential number of scattered photons is equal to the absolute value of the right side of \cref{eq:differential_photon_number}:
\begin{align}
    \diff{\Nscat} &= n\sigma\Nphoton\diff{r},\\
    &= A\Nphoton\frac{\diff{\tilde{r}}}{\tilde{r}^2},
\end{align}
where $\diff{\Nscat}$ is a number of photons scattered in a thin spherical shell of radius $\tilde{r}$ and thickness $\diff{\tilde{r}}$.
If the scattering process is Thomson scattering and incident photons are unpolarized, the scattering is isotropic.
In this case, the fraction of photons returning to the Be star is proportional to the solid angle of the star viewed from the scattering point.
Assuming the spherical star, the solid angle of the star $\Omega$ can be calculated as follows:
\begin{align}
    \Omega &= \int_0^{\sin^{-1}{\frac{1}{\tilde{r}}}}2\pi\sin{\theta}\diff{\theta},\\
    &= 2\pi\brkts{1-\sqrt{1-\tilde{r}^{-2}}},\\
    &= 2\pi F(\tilde{r}^{-1}). & \brkts{F(x)\equiv1-\sqrt{1-x^2}}
\end{align}
Then, the differential number of returning photons $\diff{\Nret}$ can be calculated from $\diff{\Nscat}$ and $\Omega$:
\begin{align}
    \diff{\Nret} &= \frac{\Omega}{4\pi}\diff{\Nscat},\\
     &= \frac{A}{2}\Nphoton F(\tilde{r}^{-1})\frac{\diff{\tilde{r}}}{\tilde{r}^2}.\label{eq:differential_return_photons}
\end{align}
We can obtain the total number of returning photons $\Nret$ by integrating $\mathrm{d}\Nret$ from the stellar surface to the infinity:
\begin{equation}
    \Nret = \int_1^\infty \frac{A}{2}\Nphoton F(\tilde{r}^{-1})\frac{\diff{\tilde{r}}}{\tilde{r}^2}.
\end{equation}
If $\Nphoton=\Nphoton_0$, the fraction of returning photons $\Nret/\Nphoton_0$ is calculated as follows:
\begin{align}
    \frac{\Nret}{\Nphoton_0} =& \frac{A}{2}\int_1^\infty F(\tilde{r}^{-1})\frac{\diff{r}}{\tilde{r}^2},\\
    =& \frac{A}{2}\int_0^1F(u)\diff{u}, & \brkts{u=\tilde{r}^{-1}}\\
    =& \frac{A}{2}\brkts{1-\frac{\pi}{4}} = 0.11A.\label{eq:return_photons1}
\end{align}
Similarly, the case $\Nphoton=\Npass$ is as follows:
\begin{align}
    \frac{\Nret}{\Nphoton_0} =& \frac{A}{2}\int_1^\infty \exp\brktl{-A\brkts{1-\tilde{r}^{-1}}}F(\tilde{r}^{-1})\frac{\diff{\tilde{r}}}{\tilde{r}^2},\\
    =& \frac{A}{2}\napier^{-A}\int_0^1F(u)\napier^{Au}\diff{u},\\
    =& \frac{A}{2}\napier^{-A}\brkts{
        \frac{\napier^{A}-1}{A} - \int_0^1\sqrt{1-u^2}\napier^{Au}\diff{u}
    }.\label{eq:return_photons2}
\end{align}
Here the first order modified Bessel and Struve function of the first kind, $I_1$ and $L_1$ can be written in following integral representations:
\begin{align}
    I_1(x) &= \frac{2}{\pi}x\int_0^1\sqrt{1-u^2}\cosh{(xu)}\diff{u},\\
    L_1(x) &= \frac{2}{\pi}x\int_0^1\sqrt{1-u^2}\sinh{(xu)}\diff{u}.
\end{align}
Hence,
\begin{align}
    \pi\frac{I_1(x) + L_1(x)}{2x} &= \int_0^1\sqrt{1-u^2}\brktl{\cosh{(xu)}+\sinh{(xu)}}\diff{u},\\
    &= \int_0^1\sqrt{1-u^2}\napier^{xu}\diff{u}.
\end{align}
Therefore,
\begin{align}
    \text{Right side of \eqref{eq:return_photons2}} &= \frac{A}{2}\napier^{-A}\brktl{
        \frac{\napier^{A}-1}{A} - \pi\frac{I_1(A) + L_1(A)}{2A}
    },\\
    &= \frac{1}{2}\brktm{
        1 - \napier^{-A}\brktl{1 + \pi\frac{I_1(A) + L_1(A)}{2}}
    }.\label{eq:return_photons3}
\end{align}
Substituting $\sigma=\sigma_\mathrm{T}=6.7\times10^{-25}~\mathrm{cm^2}$ and $N=2.4\times10^{24}~\mathrm{cm}^{-2}$ to \cref{eq:return_photons1,eq:return_photons3}, the fraction is calculated as $\Nret/\Nphoton_0=0.13, 0.17$.
In other words, about 10--20 \% of the total photons return to the Be star in the model considered in \cref{sec:scenario2}.


\bsp	
\label{lastpage}
\end{document}